\documentclass[footinbib,prb,preprint,aps]{revtex4-2}
\usepackage{graphicx}
\usepackage{bm}
\usepackage{soul}
\usepackage{bbm}
\usepackage{amsmath,amsfonts,amssymb}
\usepackage{extarrows} 
\usepackage{mathtools}
\usepackage{empheq}
\usepackage{xcolor}
\usepackage{physics}

\usepackage[colorlinks = true,
linkcolor = magenta,anchorcolor = magenta,
citecolor = magenta,filecolor = magenta,
urlcolor = magenta]{hyperref}

\usepackage{tikz}
\usepackage{orcidlink}

\usepackage[normalem]{ulem}

\definecolor{myred}{RGB}{210, 4, 45}

\newcommand{\stkout}[1]{\ifmmode\text{\sout{\ensuremath{#1}}}\else\sout{#1}\fi}

\newcommand{\XYZ}[2]{{\iffalse {#1} \fi}{#2}}

\usepackage[left=20mm,right=20mm,top=20mm,bottom=20mm]{geometry}
\usepackage[utf8]{inputenc}

\definecolor{babyblue}{rgb}{0.54, 0.81, 0.94}
\definecolor{atomictangerine}{rgb}{1.0, 0.6, 0.4}

\newcommand{\nc}{\newcommand}
\nc{\nn}{\nonumber}
\def\e{\mathcal{E}}

\def\R{\mathcal{R}}
\def\ee{\textrm{e-e}}
\def\eph{\textrm{e-ph}}
\def\eimp{\textrm{e-imp}}
\def\epht{\textrm{photon-e}}

\usepackage[colorlinks = true,
linkcolor = magenta,anchorcolor = magenta,
citecolor = magenta,filecolor = magenta,
urlcolor = magenta]{hyperref}

\begin{document}

\title{\XYZ{Comprehensive}{Quantum-optical} theory of the \XYZ{coherent}{few femtosecond} nonlinear optical response of \XYZ{electrons in}{} Drude metals with a non-parabolic conduction band}
\date{\today}

\author{Ieng-Wai Un\,\orcidlink{0000-0002-7156-0019}}
\email{iengwai@m.scnu.edu.cn}
\affiliation{Key Laboratory of Atomic and Subatomic Structure and Quantum Control (Ministry of Education), Guangdong Basic Research Center of Excellence for Structure and Fundamental Interactions of Matter, School of Physics, South China Normal University, Guangzhou 510006, China;}
\affiliation{
Guangdong Provincial Key Laboratory of Quantum Engineering and Quantum Materials, Guangdong-Hong Kong Joint Laboratory of Quantum Matter, South China Normal University, Guangzhou 510006, China.}

\author{Subhajit Sarkar\,\orcidlink{0000-0002-7260-5100}}
\affiliation{Department of Physics, School of Natural Sciences, Shiv Nadar Institution of Eminence Deemed to be University, Delhi-NCR, NH91, Tehsil Dadri, Greater Noida, Uttar Pradesh 201314, India}


\author{Yonatan Sivan\,\orcidlink{0000-0003-4361-4179}}
\affiliation{School of Electrical and Computer Engineering, Ben-Gurion University of the Negev, Beer Sheva, 8410501, Israel.}

\begin{abstract}
We develop an energy-space density matrix framework to investigate the interaction of extremely short optical pulses (ESPs) with transparent conducting oxides (TCOs). This approach captures not only electron populations, material polarization, and the permittivity, but also the the quantum coherences between states. Compared to traditional momentum-space models, the energy-space formulation offers substantial computational simplification while retaining accuracy. Building on but going beyond the scope of Ref.~\cite{single_cycle_nlty_Letter}, we focus on dynamical features previously unexplored. Our formulation reveals clear signatures of quantum coherence in the net absorption dynamics and highlights the emergence of strong excited-state absorption under intense excitation. \XYZ{It also clarifies that spontaneous emission can be neglected in this regime}{}. Furthermore, we investigate the influence of pump pulse intensity on the local field’s duration, spectral broadening and shift, and phase induced by carrier dynamics, highlighting the absorptive nature of the nonlinear response. Our results provide a unified framework for understanding nonlinear light-matter interaction in dispersive, low-density electron systems driven far from equilibrium by intense broadband excitation.
\end{abstract}

\maketitle

\section{Introduction}
The study of the interaction of an extremely short pulse with condensed matter is a topic of fundamental importance. It is required for the study of effects such as attosecond pulse and high-harmonics generation~\cite{Lu_review_2023,Meier_Koch_PRB_2008,Vampa_PRL_2014}, ionization dynamics~\cite{Nirit_2012}, metallization of dielectrics~\cite{Stockman-current-in-dielectrics,Stockman-current-in-dielectrics2,Khurgin_current_in_dielectrics}, generation of ultrashort currents~\cite{Ossiander_LiF_currents_NatComm_2022} etc.. While {\em real-space} techniques such as real-time Time-Dependent Density Functional Theory (TD-DFT) are extremely successful in modeling these interactions down to quantum and transport effects in nanometric systems~\cite{Marinica-quantum-dimer,Aizpurua-Hefei}, they are highly specialized and require very heavy computational resources. As a result, only a limited number of experts are capable of performing such numerical studies. This approach also has a limited capability to track the electron distribution dynamics in energy and time. As a result, effects of high temperature, deviation from thermal equilibrium, population inversion, and associated effects of excited state absorption and stimulated emission are usually not studied with this approach; instead, population inversion is usually treated with the much simpler incoherent rate equations.

An alternative approach relies on the Maxwell-Bloch (or Density-Matrix, DM) formulation, which tracks the electron state occupation dynamics in {\em momentum space}. Like TD-DFT, these approaches require heavy computations when applied to solid-state systems. Nevertheless, significant simplification can be achieved by transforming this formulation from momentum to energy-space, i.e., by summing over the angular degrees of freedom of the momentum. The resulting formulation 
involves only a one-dimensional formulation compared to the original three-dimensional momentum space formulation without compromising accuracy. 

Here, we use such an approach to study the interactions of a single-cycle intense optical pulse with a transparent conducting oxide (TCO), specifically, an Indium-Tin-Oxide (ITO) nanosphere, using the same material parameters as in our previous works~\cite{Un-Sarkar-Sivan-LEDD-I,Un-Sarkar-Sivan-LEDD-II}. \XYZ{}{At infrared and telecom frequencies, ITO is well described by a Drude model~\cite{Guru_Boltasseva_alternatives} due to its electron density being significantly higher than that of intrinsic semiconductors; yet, since that density is substantially lower than that of noble metals, ITO exhibits a characteristic epsilon-near-zero (ENZ) response in the near-infrared~\cite{Boyd_NLO_ENZ_ITO,Khurgin_ENZs_nlty}.} 
This material has attracted significant interest recently due to its strong nonlinear optical response~\cite{Boyd_NLO_ENZ_ITO,Kinsey-Nat-Rev-Mater-2019}, and claims for an instantaneous response~\cite{Sapienza_double_slit_2022,Shalaev_Segev_nanophotonics_2023}. \XYZ{}{Microscopically, the large nonlinearity arises from the non-parabolic conduction band~\cite{Kinsey_ENZ_OptMatExp} and from the strong field dependence of the electron–electron and electron–phonon collision rates~\cite{Un-Harcavi-Sivan-LEDD-Viewpoint}.} Our approach adapts ideas originally developed to deal with CW to few 10's of femtosecond long pulses interacting with metals and TCOs~\cite{delFatti_nonequilib_2000,GdA_hot_es,Dubi-Sivan,Sarkar-Un-Sivan-Dubi-NESS-SC,Un-Sarkar-Sivan-LEDD-I,Un-Sarkar-Sivan-LEDD-II}, but extends them to account not only for electron population and permittivity dynamics, but also to coherences between the electron states (i.e., the off-diagonal elements of the density matrix). 

Initial results on this problem were published in~\cite{single_cycle_nlty_Letter}, which focused on the characterization of the {\em strength} of the nonlinear response (in terms of the electron excitation dynamics, material polarization and permittivity). Instead, here, we focus on the {\em formulational aspects} of the problem, as well as on additional aspects of the dynamics that were not considered in~\cite{single_cycle_nlty_Letter}. Specifically, in Section~\ref{sec:formulation}, we provide a detailed derivation of the DM formulation that underlies~\cite{single_cycle_nlty_Letter}, including the transition dipole matrix element calculation in momentum space, and its transformation to the far more computationally-economic energy-space. Then, in Section~\ref{sec:Results}, we perform a thorough comparison of the strengths of the various terms appearing in the DM equations and analyze the dynamics of the coherence between the various electron states as well as the resulting repeated frequency doubling dynamics. Then, we identify the signature of the coherence in the (net) energy-integrated absorption dynamics and the emergence of significant excited state absorption within its time-integrated energy spectra; the latter is responsible for the super-linear nature of the absorption, hence, counteracting the effects of population inversion. We also justify the neglect of spontaneous emission (see Appendix.~\ref{app:PL}). 

Then, we characterize systematically the effects of the pump pulse intensity on the local field duration, spectrum (broadening and shift), and phase; we identify features that point to the absorptive nature of the nonlinear response, and can be measured in experiment. In Section~\ref{sec:conclusion}, we conclude with a short discussion on the ways to connect our approach to similar studies of short-pulse interactions with dielectrics and semiconductors.

\section{Formulation}\label{sec:formulation}
\subsection{A derivation of a momentum-conserving energy-space density matrix formulation}

Consider a many-electron system interacting with an electromagnetic field $E(t)$ and a phonon bath. The total Hamiltonian can be written as 
\begin{align}\label{eq:H_tot_e_phn}
\hat{\mathcal{H}}_\textrm{tot} = \hat{\mathcal{H}}_\textrm{e,0} + \hat{\mathcal{H}}_\textrm{ph,0} + \hat{\mathcal{H}}_\epht + \hat{\mathcal{H}}_\ee + \hat{\mathcal{H}}_\eph + \hat{\mathcal{H}}_\textrm{e-imp}.
\end{align}
Here, $\hat{\mathcal{H}}_\textrm{e,0}$ is the single-particle approximation for the Hamiltonian of the electrons in the lattice periodic potential, $\hat{\mathcal{H}}_\textrm{ph,0}$ is the unperturbed Hamiltonian of the phonon subsystem, and $\mathcal{H}_\epht$ is the electric dipole interaction Hamiltonian 
\begin{align}\label{eq:H_int}
\hat{\mathcal{H}}_\epht = - \XYZ{\hat{d}}{\hat{d}_z}
\cdot E(t),
\end{align}
\XYZ{with $\hat{d}$ being the electric transition dipole operator and $E(t)$ being the local electric field, calculated self-consistently with the density matrix elements, see Section~\ref{sub:Esca_Eint_NP_TD}.}{where $E(t)$ is the local electric field, determined self-consistently from the density-matrix elements (see Section~\ref{sub:Esca_Eint_NP_TD}). Without loss of generality, the electric field is assumed to be in the $z$-direction. $\hat{d}_z$ is the electric transition dipole operator along the direction of the electric field.} In that regard, we neglect electron- and phonon-assisted absorption events as well as interband transitions in favor of Landau-damping-based absorption (see discussion of Ref.~\cite{Khurgin-Faraday-hot-es} on the analysis of Ref.~\cite{hot_es_Atwater}). Additionally, we neglect the contribution of spontaneous emission (photoluminescence), see justification in Appendix.~\ref{app:PL}. $\hat{\mathcal{H}}_\ee$ is the electron-electron interaction via the Coulomb potential, $\hat{\mathcal{H}}_\eph$ is the electron-phonon interaction via the deformation potential and $\hat{\mathcal{H}}_\eimp$ is the electron-impurity interaction. 

The evolution of the total system is given by the von Neumann equation
\begin{align}\label{eq:vN_tot}
\dot{\rho}_\textrm{tot} = -\dfrac{i}{\hbar}\left[ 
\hat{\mathcal{H}}_\textrm{tot}, {\rho}_\textrm{tot} \right],
\end{align}
where ${\rho}_\textrm{tot}$ is the density matrix of the total system. Since we are mostly interested in the dynamics of the electron subsystem, we follow the procedure shown in~\cite{Sarkar-Un-Sivan-DM,Bustamante_JCP} to infer the equations of motion of the electron subsystem (so-called reduced equations of motion) from the equation of motion of the total system~\eqref{eq:vN_tot}. This gives us (see more details in Refs.~\cite{intro_Lindblad_eq,Breuer,intro_open_qm}) 
\begin{align}\label{eq:vN_elec}
\dot{\rho}(t) = - \dfrac{i}{\hbar}\left[\hat{\mathcal{H}}_e,\rho\right] + \R(\rho) = - \dfrac{i}{\hbar}\left[\hat{\mathcal{H}}_0 + \hat{\mathcal{H}}_\epht,\rho\right] + \R(\rho),
\end{align}
where $\rho$ is the (reduced) density matrix of the electron subsystem whose diagonal elements are equal to the population (i.e., $\rho_{{\bf k}{\bf k}}(t) = f_{{\bf k}}(t)$) and $\hat{\mathcal{H}}_0 = \hat{\mathcal{H}}_{e,0} + \hat{\mathcal{H}}_\textrm{L,e-e} + \hat{\mathcal{H}}_\textrm{L,e-ph} + \hat{\mathcal{H}}_\textrm{L,e-imp}$ represents the effective Hamiltonian of the electron subsystem, derived from the full many-body Hamiltonian~\eqref{eq:H_tot_e_phn}; it includes the real part of the self-energy resulting from $\ee$, $\eph$, and $\eimp$ interactions~\cite{Quantum-Liquid-Coleman,Breuer}, which constitute the so-called Lamb-shift Hamiltonian~\cite{Breuer,intro_Lindblad_eq} whose effect is to renormalize the electron subsystem energy levels, resulting in a modification of the band dispersion and the electron density of states. $\R$ is the relaxation operator~\cite{Blum_DM_book} originating from the imaginary part of the self-energy~\cite{Quantum-Liquid-Coleman}, responsible for the relaxation dynamics resulting from the $\ee$ and $\eimp$ interactions and for the dissipative dynamics resulting from $\eph$ interactions. Under these conditions, Eq.~\eqref{eq:vN_elec} is equivalent to
\begin{subequations}\label{eq:EOM_rho}
\begin{empheq}[left=\empheqlbrace]{align}
&\begin{aligned}
\dot{f}_{{\bf k}} &= \R_{{\bf k}{\bf k}}(\rho(t)) + \dfrac{i}{\hbar}E(t)\sum_{{\bf k}'}\left[\XYZ{d_{{\bf k}' {\bf k}}}{(d_z)_{{\bf k}' {\bf k}}}\rho_{{\bf k}{\bf k}'} - \XYZ{d_{{\bf k}{\bf k}'}}{(d_z)_{{\bf k}{\bf k}'}}\rho_{{\bf k}'{\bf k}} \right],
\end{aligned}\label{eq:EOM_rho_aa} \\
&\begin{aligned}
\dot{\rho}_{{\bf k}{\bf k}'} 
= \R_{{\bf k}{\bf k}'}(\rho(t)) - \dfrac{i}{\hbar}\left(\e_{{\bf k}} - \e_{{\bf k}'}\right) \rho_{{\bf k}{\bf k}'} - \dfrac{i}{\hbar} E(t) \XYZ{d_{{\bf k}{\bf k}'}}{(d_z)_{{\bf k}{\bf k}'}} \left(f_{{\bf k}}(t) - f_{{\bf k}'}(t)\right) 
\\+ \dfrac{i}{\hbar} E(t) \sum_{{\bf k}'' \neq {\bf k},{\bf k}'} \left[\XYZ{d_{{\bf k}{\bf k}''}}{(d_z)_{{\bf k}{\bf k}''}} \rho_{{\bf k}''{\bf k}'} - \rho_{{\bf k}{\bf k}''}\XYZ{d_{{\bf k}''{\bf k}'}}{(d_z)_{{\bf k}''{\bf k}'}}\right]
\end{aligned} & \text{for }{\bf k} \neq {\bf k}', & \label{eq:EOM_rho_ab}
\end{empheq}
\end{subequations}
where $\omega_{{\bf k}{\bf k}'} \equiv \left(\e_{{\bf k}} - \e_{{\bf k}'}\right)/\hbar$ with $\e_{{\bf k}}$ and $\e_{{\bf k}'}$ being the renormalized energy of individual electrons and \XYZ{$d_{{\bf k}{\bf k}'}$}{$(d_z)_{{\bf k}{\bf k}'}$} are the matrix elements of the electric dipole transition operator \XYZ{}{along the direction of the electric field}. 


We determine the momentum space dipole transition matrix elements \XYZ{$d_{{\bf k}{\bf k}'}$}{$(d_z)_{{\bf k}{\bf k}'}$} from the matrix elements of the momentum operator \XYZ{$p_{{\bf k}{\bf k}'}$}{$(p_z)_{{\bf k}{\bf k}'}$ and $(p^2)_{{\bf k}{\bf k}'} = (p_x^2+p_y^2+p_z^2)_{{\bf k}{\bf k}'}$} using the relation
\begin{align}\label{eq:p2d}
\XYZ{(d_{x,y,z})_{{\bf k}{\bf k}'}}{(d_z)_{{\bf k}{\bf k}'}} = - \dfrac{2ie\hbar}{(p^2)_{{\bf k}{\bf k}}-(p^2)_{{\bf k}'{\bf k}'}} \XYZ{(p_{x,y,z})_{{\bf k}{\bf k}'}}{(p_z)_{{\bf k}{\bf k}'}},
\end{align}
where $e$ is the electron charge. Eq.~\eqref{eq:p2d} is derived from the commutation relation between the position operator and the operator $\hat{{\bf p}}^2$ (see~\cite[Ch. 5]{Haug_Koch_book}), 
while the matrix elements of the momentum operator \XYZ{$p_{{\bf k}{\bf k}'}$}{$\left(p_{x,y,z}\right)_{{\bf k}{\bf k}'}$} are determined by assuming an infinitely deep 1D potential well representing the sample geometry, as in popular earlier works~\cite{Govorov_ACS_phot_2017,Khurgin-Levy-ACS-Photonics-2020}. 

The simplest phenomenological form of the relaxation operator $\R$ can be expressed as~\cite{fain_QM_elec_book}
\begin{subequations}\label{eq:relax_op_phnm}
\begin{empheq}[left=\empheqlbrace]{align}
\R_{{\bf k}{\bf k}}(t) &= \left(\dfrac{\partial f_{{\bf k}}}{\partial t}\right)_\textrm{e-e} + \left(\dfrac{\partial f_{{\bf k}}}{\partial t}\right)_\textrm{e -ph} + \left(\dfrac{\partial f_{{\bf k}}}{\partial t}\right)_\textrm{e -imp} & \label{eq:relax_op_aa} \\
\R_{{\bf k}{\bf k}'}(t) &= -\eta_{{\bf k}{\bf k}'}(t)\rho_{{\bf k}{\bf k}'}(t) &\textrm{ for } {\bf k} \neq {\bf k}'. \label{eq:relax_op_phnm_ab}
\end{empheq}
\end{subequations}
Here, the diagonal terms $\R_{{\bf k}{\bf k}}$ are responsible for the relaxation of the conduction electrons; they are treated the same way as in Refs.~\cite{Un-Sarkar-Sivan-LEDD-I,Un-Sarkar-Sivan-LEDD-II} using the Fermi’s golden rule as employed in Ref.~\cite{Snoke_solid_state}, including also the temperature dependence of the Thomas-Fermi momentum Eq.~(15) of Ref.~\cite{Un-Sarkar-Sivan-LEDD-I}~\footnote{Note that the factor $4\pi^3$ in that equation should be removed.}. On the other hand, $\eta_{{\bf k}{\bf k}'}(t) = (\eta_{{\bf k}}(t) + \eta_{{\bf k}'}(t))/2$ is the decoherence rate between states $\ket{{\bf k}}$ and $\ket{{\bf k}'}$ for ${\bf k}\neq{\bf k}'$~\cite{Bustamante_JCP,Sarkar-Un-Sivan-DM}, and $\eta_{{\bf k}} = \delta \R_{{\bf k}{\bf k}}/\delta f_{{\bf k}}$ is the functional derivative of $\R_{{\bf k}{\bf k}}$ with respect to $f_{{\bf k}}$. This phenomenological model corresponds to replacing $\partial_t$ by $\partial_t +\eta_{{\bf k}{\bf k}'}$ (see the replacement in Eq.~\eqref{eq:EOM_rho_ab}). 
Put simply, Eq.~\eqref{eq:relax_op_phnm} amounts to computing the relaxation of the diagonal terms consistently, and using the relaxation time approximation (RTA) for the off-diagonal terms.

Unfortunately, as shown in Ref.~\cite{relax_op_qp_2020}, using Eq.~\eqref{eq:relax_op_phnm_ab} in the equations of motion Eq.~\eqref{eq:EOM_rho} leads to a violation of the charge density continuity equation~\footnote{\XYZ{}{Multiplying Eq.~\eqref{eq:vN_elec} by the dipole moment operator $\hat{{\bf d}} = (-e)\hat{{\bf r}}$, taking the trace, and using the relation between the velocity and position operator $\hat{{\bf v}} = (i/\hbar)[\hat{\mathcal{H}}_e,\hat{{\bf r}}]$, we obtained $\partial {\bf P}/\partial t = {\bf J} + \mathrm{Tr}\left( \hat{{\bf d}} \mathcal{R}(\rho) \right)/V$, where ${\bf P} = (1/V)\mathrm{Tr}\left( \hat{{\bf d}} \hat{\rho} \right)$ is the polarization density and ${\bf J} = (1/V)\mathrm{Tr}\left( (-e)\hat{{\bf v}} \hat{\rho} \right)$ is the (polarization) current density. For the relaxation operator given by Eq.~\eqref{eq:relax_op_phnm_ab}, the term $\mathrm{Tr}\left( \hat{{\bf d}} \hat{\mathcal{R}} \right)$ is, in general, nonzero. Consequently, the (polarization) current density is not equal to the temporal derivative of the current density if $\mathrm{Tr}\left( \hat{{\bf d}} \hat{\mathcal{R}} \right) \neq 0$, leading to a violation of the charge density continuity equation.}}, as well as to an incorrect conductivity in the static limit~\cite{corr_to_relax_op,Mermin_Lindhard_RTA}. These inconsistencies can be resolved by constructing the relaxation operator so that it satisfies $\textrm{Tr}\left(\hat{{\bf d}} \hat{\R}\right) = 0$ (see details in Ref.~\cite{corr_to_relax_op,relax_op_qp_2020}). Following the procedure shown in Refs.~\cite{corr_to_relax_op,relax_op_qp_2020}, a relatively simple relaxation operator preserving the continuity equation can be obtained by replacing Eq.~\eqref{eq:relax_op_phnm_ab} with~\footnote{\XYZ{}{For the relaxation operator given by Eq.~\eqref{eq:relax_op_ab}, $\mathrm{Tr}\left( \hat{{\bf d}} \hat{\mathcal{R}} \right) = \sum_{{\bf k},{\bf k}'}d_{{\bf k}'{\bf k}}\mathcal{R}_{{\bf k}{\bf k}'} = -\sum_{{\bf k},{\bf k}'}\eta_{{\bf k}{\bf k}'}\left[ d_{{\bf k}'{\bf k}}\rho_{{\bf k}{\bf k}'} - d_{{\bf k}{\bf k}'}\rho_{{\bf k}'{\bf k}} \right] = 0$.}}
\begin{align}\label{eq:relax_op_ab}
\R_{{\bf k}{\bf k}'}(t) &= - \eta_{{\bf k}{\bf k}'}(t)(\rho_{{\bf k}{\bf k}'}(t) - h_{{\bf k}{\bf k}'} \rho_{{\bf k}'{\bf k}}(t)), & \textrm{ for }{\bf k} \neq {\bf k}',
\end{align}
where $h_{{\bf k}{\bf k}'}$ is a unitless phase factor such that \XYZ{$h_{{\bf k}{\bf k}'} d_{{\bf k}'{\bf k}} = d_{{\bf k}{\bf k}'}$}{$h_{{\bf k}{\bf k}'} (d_z)_{{\bf k}'{\bf k}} = (d_z)_{{\bf k}{\bf k}'}$}~\footnote{In general, $(d_z)_{{\bf k}'{\bf k}}$ and $(d_z)_{{\bf k}{\bf k}'}$ differ by a phase factor. For our choice of a 1D symmetric potential, we have $(d_z)_{{\bf k}'{\bf k}} = (d_z)_{{\bf k}{\bf k}'}$.}.

We further assume that the electric field $E$ is weak enough so that coherent optical nonlinearities are negligible. These effects are generally weak, especially since they do not enjoy the resonant enhancement experienced by the photon absorption events we do account for. Indeed, the efficiency in the measurements reported in Ref.~\cite{Yang-HHG-CdO-natphys-2019} for such effects was $10^{-5}$ for 3rd harmonic generation, and $10^{-8}$ and $10^{-10}$ for 5th and 7th harmonics, respectively. In Ref.~\cite{Korobenko-HHG-TiN-natcommun-2021}, harmonics were obtained only for TiN, and with intensities 10 times higher than employed here and ever for ITO. This allows neglecting the last term in Eq.~\eqref{eq:EOM_rho_ab} (see Ref.~\cite{Boyd-book}) which thus becomes
\begin{align}\label{eq:rho_ab_dot}
\dot{\rho}_{{\bf k}{\bf k}'} 
= \R_{{\bf k}{\bf k}'}(\rho(t)) - \dfrac{i}{\hbar}\left(\e_{{\bf k}} - \e_{{\bf k}'}\right) \rho_{{\bf k}{\bf k}'} - \dfrac{i}{\hbar} E(t) \XYZ{d_{{\bf k}{\bf k}'}}{ (d_z)_{{\bf k}{\bf k}'} } \left(f_{{\bf k}}(t) - f_{{\bf k}'}(t)\right).
\end{align}
We then substitute Eq.~\eqref{eq:relax_op_aa} into Eq.~\eqref{eq:EOM_rho_aa}, and substitute Eq.~\eqref{eq:relax_op_ab} into Eq.~\eqref{eq:rho_ab_dot}. This gives
\begin{subequations}\label{eq:BE_rho_EOM_k}
\begin{empheq}[left=\empheqlbrace]{align}
&\begin{aligned}
\dfrac{\partial f_{{\bf k}}(t)}{\partial t} &= \left(\dfrac{\partial f_{{\bf k}}}{\partial t}\right)_{\ee} + \left(\dfrac{\partial f_{{\bf k}}}{\partial t}\right)_{\eph} + \left(\dfrac{\partial f_{{\bf k}}}{\partial t}\right)_{\eimp} \\ &\qquad\qquad\qquad - \dfrac{i}{\hbar}E(t)
\sum_{{\bf k}'}
\left[\XYZ{d_{{\bf k}'{\bf k}}}{(d_z)_{{\bf k}'{\bf k}}}\rho_{{\bf k}{\bf k}'}(t) - \XYZ{d_{{\bf k}{\bf k}'}}{(d_z)_{{\bf k}{\bf k}'}}\rho_{{\bf k}'{\bf k}}(t) \right]
\end{aligned}\label{eq:BE_EOM_k} \\
&\begin{aligned}
\dfrac{\partial\rho_{{\bf k}{\bf k}'}(t)}{\partial t} = - \eta_{{\bf k}{\bf k}'}\left( \rho_{{\bf k}{\bf k}'} - h_{{\bf k}{\bf k}'} \rho_{{\bf k}'{\bf k}} \right) - \dfrac{i}{\hbar}\left(\e_{{\bf k}} - \e_{{\bf k}'}\right)\rho_{{\bf k}{\bf k}'}
- \dfrac{i}{\hbar}E(t)\XYZ{d_{{\bf k}{\bf k}'}}{(d_z)_{{\bf k}{\bf k}'}}\left[ f_{{\bf k}}(t) - f_{{\bf k}'}(t)\right],
\end{aligned}\label{eq:rho_EOM_k}
\end{empheq}
\end{subequations}
The polarization density \XYZ{}{along the direction of the electric field} can be obtained by taking the trace of the matrix multiplication of the density matrix and the dipole transition matrix, namely,
\begin{align}\label{eq:P_k}
P(t) = \dfrac{2}{V}\sum_{{\bf k},{\bf k}'}\XYZ{d_{{\bf k}'{\bf k}}}{(d_z)_{{\bf k}'{\bf k}}}\rho_{{\bf k}{\bf k}'}(t).
\end{align}

\XYZ{}{The initial condition for the resulting system of equation includes diagonal terms at room temperature thermal distribution for the electrons (and for phonons as well), and vanishing off-diagonal terms.}

\subsection{From $k$ to $\e$ space formulation}

We now would like to convert Eq.~\eqref{eq:BE_rho_EOM_k} and Eq.~\eqref{eq:P_k} to energy space; this has the advantage of significantly shortening the run times and reducing memory consumption. To do that, we first replace $\displaystyle\sum_{{\bf k}'}$ in~ Eq.~\eqref{eq:BE_EOM_k} with~\footnote{Since the optical transition is spin-conserving, we do not need to introduce a factor of 2 for the electron spin when summing over all ${\bf k}$-states.} 
$$
\dfrac{V}{(2\pi)^3} \displaystyle \int d^3 k' = \dfrac{V}{(2\pi)^3} \int \pi^2 \rho_e(\e')d\e'd\Omega_{{\bf k}'}.
$$ 
with $\rho_e(\e) = ({k^2}/{\pi^2})({dk}/{d\e})$ being the electron density of states (eDOS) and $V$ the particle volume. \XYZ{}{The energy–momentum relation $k(\e)$ is taken from the Kane quasilinear dispersion~\cite{Kane-quasilinear}, accounting for conduction-band nonparabolicity in ITO~\cite{Guo_ITO_nanorod_natphoton,ITO_Nonparabolicity},
\begin{align}
\hbar k(\mathcal{E}) = \sqrt{2m_e^\ast \mathcal{E}(1+C\mathcal{E})},
\end{align}
where $m_e^\ast = 0.3964\ m_e$ is the electron effective mass at the conduction-band minimum and $C = 0.4191 \ \textrm{eV}^{-1}$ is the first-order nonparabolicity factor~\cite{Liu_ITO_2014}.}

Then, we integrate Eq.~\eqref{eq:BE_EOM_k} over the solid angle of ${\bf k}'$ (normalized by $4\pi$) to get $$\dfrac{\partial f(\e,t)}{\partial t} = \displaystyle\int \dfrac{d\Omega_{\bf k}}{4\pi}\dfrac{\partial f_{{\bf k}}(t)}{\partial t},$$ 
and multiply Eq.~\eqref{eq:rho_EOM_k} by \XYZ{$d_{{\bf k}'{\bf k}}$}{$(d_z)_{{\bf k}'{\bf k}}$} and integrate over the solid angles of ${\bf k}$ and of ${\bf k}'$~\footnote{\XYZ{}{Note that averaging over the solid angles of ${\bf k}$ and ${\bf k}'$ is fundamentally different from stochastic averaging that destroys quantum coherence. The angular integration merely accounts for isotropic contributions in momentum space and does not suppress phase coherence. Specifically, multiplying Eq.~\eqref{eq:rho_EOM_k} by $(d_z)_{{\bf k}'{\bf k}}$ yields the equation of motion for $(d_z)_{{\bf k}'{\bf k}} \rho_{{\bf k}{\bf k}'}$, i.e., the $({\bf k},{\bf k}')$ component of the polarization density. 
Accordingly, integration over the solid angles of ${\bf k}$ and ${\bf k}'$ evaluates the total contribution to the polarization density from transitions between states with momenta $|\hbar{\bf k}|$ and $|\hbar{\bf k}'|$. Since the source term of the equation of motion for $(d_z)_{{\bf k}'{\bf k}} \rho_{{\bf k}{\bf k}'}$ is proportional to $E(t)|(d_z)_{{\bf k}{\bf k}'}|^2(f_{{\bf k}} - f_{{\bf k}'}) = E(t)|(d_z)_{{\bf k}{\bf k}'}|^2\big[f(\e(|{\bf k}|)) - f(\e(|{\bf k}'|)) \big]$, which has a definite sign (in phase), $\displaystyle\int d\Omega_{{\bf k}}d\Omega_{{\bf k}'} d_{{\bf k}'{\bf k}} \rho_{{\bf k}{\bf k}'} $ produces a finite contribution to the polarization density.}} (see details in Ref.~\cite{Sarkar-Un-Sivan-DM}). The photon-e interaction term, the off-diagonal density matrix equation of motion Eq.~\eqref{eq:rho_EOM_k}, and the polarization density Eq.~\eqref{eq:P_k} thus, respectively, become
\begin{align}\label{eq:dfdt_exc_E}
\left(\dfrac{\partial f(\e,t)}{\partial t}\right)_{\epht} = - \dfrac{i V}{2\hbar} E(t) \displaystyle \int \rho_e(\e')(\overline{d_{\e'\e} \rho_{\e\e'}} - \overline{d_{\e\e'} \rho_{\e'\e}}) d\e',
\end{align}
\begin{multline}\label{eq:drhoEE_dt}
\dfrac{\partial (\overline{d_{\e'\e} \rho_{\e\e'}})}{\partial t}  + \eta_{\e\e'} (\overline{d_{\e'\e} \rho_{\e\e'}} - \overline{d_{\e\e'}\rho_{\e'\e}}) + i\left(\dfrac{\e - \e'}{\hbar}\right)(\overline{d_{\e'\e}\rho_{\e\e'}}) \\
= -\dfrac{i}{\hbar} |D(k(\e),k(\e'),L)|^2 E(t)\left[f(\e,t) - f(\e',t)\right],
\end{multline}
\begin{align}\label{eq:Pnp_Esp}
P(t) = \dfrac{V}{4} \displaystyle \int \rho_e(\e)\rho_e(\e') (\overline{d_{\e'\e}\rho_{\e\e'}} + \overline{d_{\e\e'}\rho_{\e'\e}}) d\e d\e',
\end{align}
with $\overline{d_{\e'\e} \rho_{\e\e'}} \equiv \dfrac{1}{(4\pi)^2} \displaystyle \int d\Omega_{{\bf k}} d\Omega_{{\bf k}'} \XYZ{d_{{\bf k}'{\bf k}}}{(d_z)_{{\bf k}'{\bf k}}} \rho_{{\bf k}{\bf k}'}$ ($\e \neq \e'$) being the (angular average of the) dipole-matrix-element-weighted off-diagonal elements. Here, in analogy to the momentum-space expression, $\eta_{\e\e'} = (\eta_{\e} + \eta_{\e'})/2$ is the decoherence rate for $\e \neq \e'$. By Matthiessen’s rule~\cite{Ashcroft-Mermin}, $\eta_\e = \eta_{\e,\ee} + \eta_{\e,\eph} + \eta_{\e,\eimp}$, where $\eta_{\e,\ee}$, $\eta_{\e,\eph}$ and $\eta_{\e,\eimp}$ represent the collision rates associated with the $\ee$, $\eph$ and $\eimp$ interactions, respectively (\footnote{\XYZ{}{Here, the $\ee$ interaction in Eq.~\eqref{eq:BE_EOM_k} includes only the normal processes when determining the electron distribution. Upon angular averaging, the contribution from these normal scattering processes vanishes because of overall momentum conservation. Instead, Umklapp processes contribute to the relaxation of the polarization (or equivalently, the induced current). 
For ITO, this ratio is presently unknown; however, it is likely to be of order unity, $\mathcal{O}(1)$~\cite{Kaveh-Wiser}.}}, see also details in Refs.~\cite{Un-Sarkar-Sivan-LEDD-I,Un-Sarkar-Sivan-LEDD-II}; they are obtained via functional derivation of their corresponding interaction terms with respect to $f(\e)$ in energy space~\footnote{Note that the expression in~\cite{Un-Sarkar-Sivan-LEDD-II}, namely, Eq.~(B6), had a small typo; the expression in the rectangular brackets on the 2nd line should have read $\left(1 + \frac{q^2_{TF}}{2k^2}\right)^2 - 1$.}). $|D(k,k',L)|^2$ is the angular average of the squared dipole matrix element \XYZ{$|d_{{\bf k}{\bf k}'}|^2$}{$|(d_z)_{{\bf k}{\bf k}'}|^2$}~\eqref{eq:p2d}, i.e., 
\begin{multline}\label{eq:d2D}
|D(k,k',L)|^2 \equiv \dfrac{1}{(4\pi)^2}\int d\Omega_{{\bf k}}d\Omega_{{\bf k}'}|\XYZ{d_{{\bf k}{\bf k}'}}{(d_z)_{{\bf k}{\bf k}'}}|^2\\
= \dfrac{2e^2(2\pi L)^2}{(k L)(k' L)(k^2 - k'^2)^2 L V} \Bigg\{\dfrac{((kL)^2 - (k'L)^2)^2 - 16}{4}\ln \left(\dfrac{4 + |(kL)^2 - (k'L)^2|}{4 + (kL - k'L)^2}\right) \\
- \left[((kL)^2-(k'L)^2)^2 + 1\right] \ln \left(\dfrac{1 + |(kL)^2 - (k'L)^2|}{1 + (kL - k'L)^2}\right) \\
- \dfrac{3}{4}\left[((kL) - (k'L))^2\right]^2 \ln\left|\dfrac{k - k'}{k + k'}\right|
\Bigg\}.
\end{multline}
where $k = |{\bf k}|$ and $k' = |{\bf k}'|$ are the magnitudes of the electron momentum, and $L$ is the system size (such as particle size or thin film thickness) in the direction of the electric field $E$. $D(k(\e),k(\e'),L)$ has units of [C m], and thus, can be viewed as the (absolute value of the) dipole matrix element in energy space. While the use of a one-dimensional infinitely deep potential well to evaluate the dipole matrix elements is a simplified model, it provides a reasonable approximation for NPs larger than a few nanometers. Indeed, in such systems, the energy spacing between adjacent states is typically $\lesssim 0.02-0.05$ eV, comparable with the level width ($\eta$) and much smaller than the conduction bandwidth ($\gtrsim 4$ eV). This ensures a large number of accessible energy levels, justifying the approximation of discrete levels by a quasi-continuum and rendering the dipole matrix elements effectively insensitive to particle shape. Moreover, the angular averaging in Eq.~\eqref{eq:d2D} further mitigates the limitations of this approximation and captures the essential qualitative features of the matrix elements. Furthermore, the main effect of the exaggerated depth of the potential well is to exclude photoionization processes involving states above the vacuum level, which are anyhow not the focus of this study. A more accurate calculation of the transition dipole matrix elements can be performed using the approach in Ref.~\cite{GdA_hot_es}.

\begin{figure}[h]
\centering
\includegraphics[width=0.8\textwidth]{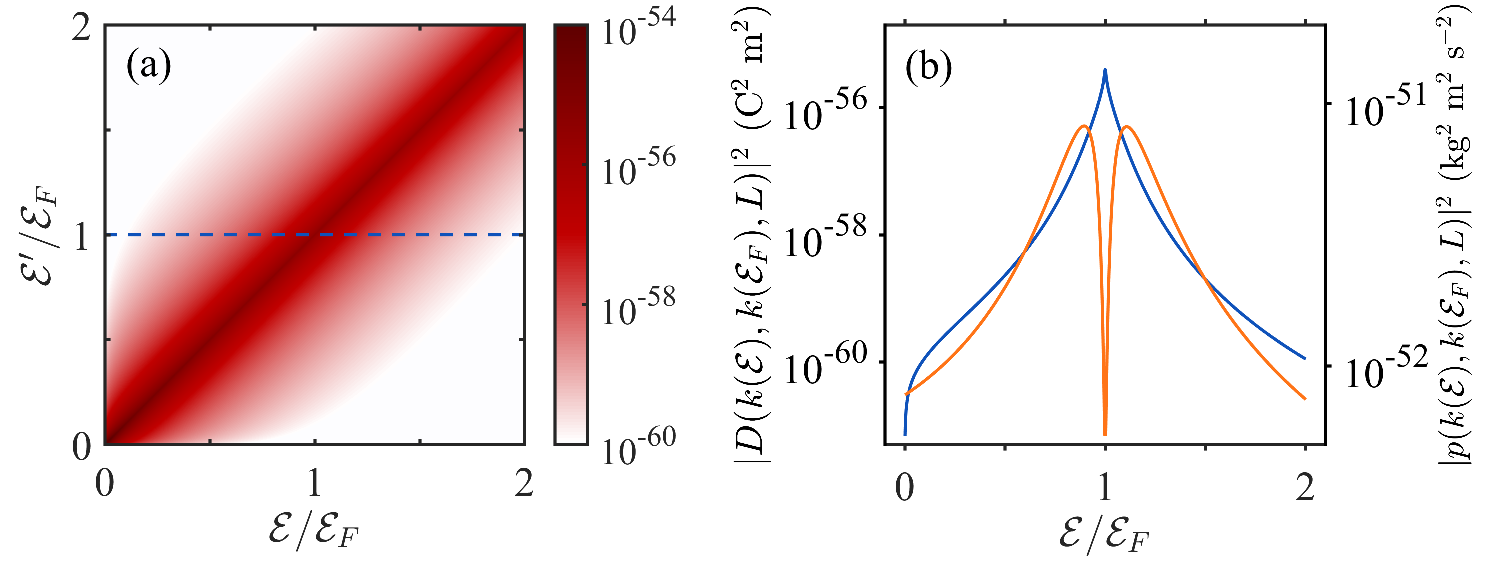}
\caption{(a) The square of the absolute value of (the angular average of) the energy space transition dipole matrix elements $|D(k(\e),k(\e'),L)|^2$ [C$^2$ m$^2$] calculated using Eq.~\eqref{eq:d2D} for $L = 4$ nm. (b) A cross-section of $|D(k(\e),k(\e'),L)|^2$ (blue solid line, labeled by the blue dashed line in (a)) and $|p(k(\e),k(\e'),L)|^2$ (orange solid line) for $\e' = \e_F$. 
}
\label{fig:mat_ele_eng}
\end{figure}

Fig.~\ref{fig:mat_ele_eng}(a) plots the angular average of the squared dipole matrix element $|D(k(\e),k(\e'),L)|^2$, showing that it is dominated by elements near the diagonal, manifesting momentum conservation in the optical transition. Fig.~\ref{fig:mat_ele_eng}(b) plots the $\e$-dependence of both $|D(k(\e),k(\e'),L)|^2$ and its corresponding momentum matrix element $|p(k(\e),k(\e'),L)|^2 = [p^2(\e) - p^2(\e')]^2 |D(k(\e),k(\e'),L)|^2 / (4e^2\hbar^2)$ (see Eq.~\eqref{eq:p2d}) at $\e' = \e_F$. One observes a divergence in $|D(k(\e),k(\e'),L)|^2$ at $\e = \e'$, but this divergence is only logarithmic. When multiplied by $(f(\e,t) - f(\e',t))$ in Eq.~\eqref{eq:drhoEE_dt}, the resulting product remains well-behaved and does not introduce any singularity or instability in the dynamics of the off-diagonal terms. In contrast, while $|p(k(\e),k(\e'),L)|^2$ is also strongly peaked near the diagonal, it vanishes exactly at $\e = \e'$, consistent with the fact that optical transitions between identical initial and final states are forbidden.

Unlike most energy space formulations (e.g., including our earlier work that includes an ad hoc normalization using Poynting's theorem~\cite{Seidman-Nitzan-non-thermal-population-model, Dubi-Sivan,Un-Sarkar-Sivan-LEDD-II}), the current approach has the advantage of incorporating both momentum and energy conservation and allowing for obtaining the photon-e interaction term and polarization density in a self-consistent manner (see details in Ref.~\cite{Sarkar-Un-Sivan-DM}). Moreover, in contrast to the heavy momentum-space discrete formulation (see e.g. Refs.~\cite{GdA_hot_es,Govorov_ACS_phot_2017}), the energy space formulation developed in this work 
involves a significantly reduced computational effort. Indeed, despite the large number of off-diagonal elements that still need to be computed, it remains significantly smaller than in the momentum space formulation~\footnote{In a direct momentum-space formulation with $N_k$ grid points per dimension, the electronic states are sampled on a three-dimensional grid of size $N_k^3$, and the evaluation of the electron–electron scattering term scales as $\mathcal{O}(N_k^6)$, in addition to requiring numerical enforcement of energy- and momentum-conservation constraints. By contrast, after transforming to energy space with $N_E$ grid points, the same term scales as $\mathcal{O}(N_E^2)$, with momentum conservation imposed analytically. This reduction leads to orders-of-magnitude savings in computational cost and memory.}. Consequently, our approach offers substantial computational savings over momentum space calculations without loss of information. This provides an advantage compared to TD-DFT (e.g., Refs.~\cite{Marinica-quantum-dimer,RT-TDDFT,Yabana_ACS_Photonics_2019}), in addition to the treatment of non-zero electron temperatures and the somewhat more systematic treatment of various electron collision mechanisms (both effects, admittedly, of low importance in the current context).

\subsection{Calculation of internal fields of a single small nanosphere in the time domain}\label{sub:Esca_Eint_NP_TD}
\begin{figure}[ht]
\centering
\includegraphics[width=0.45\linewidth]{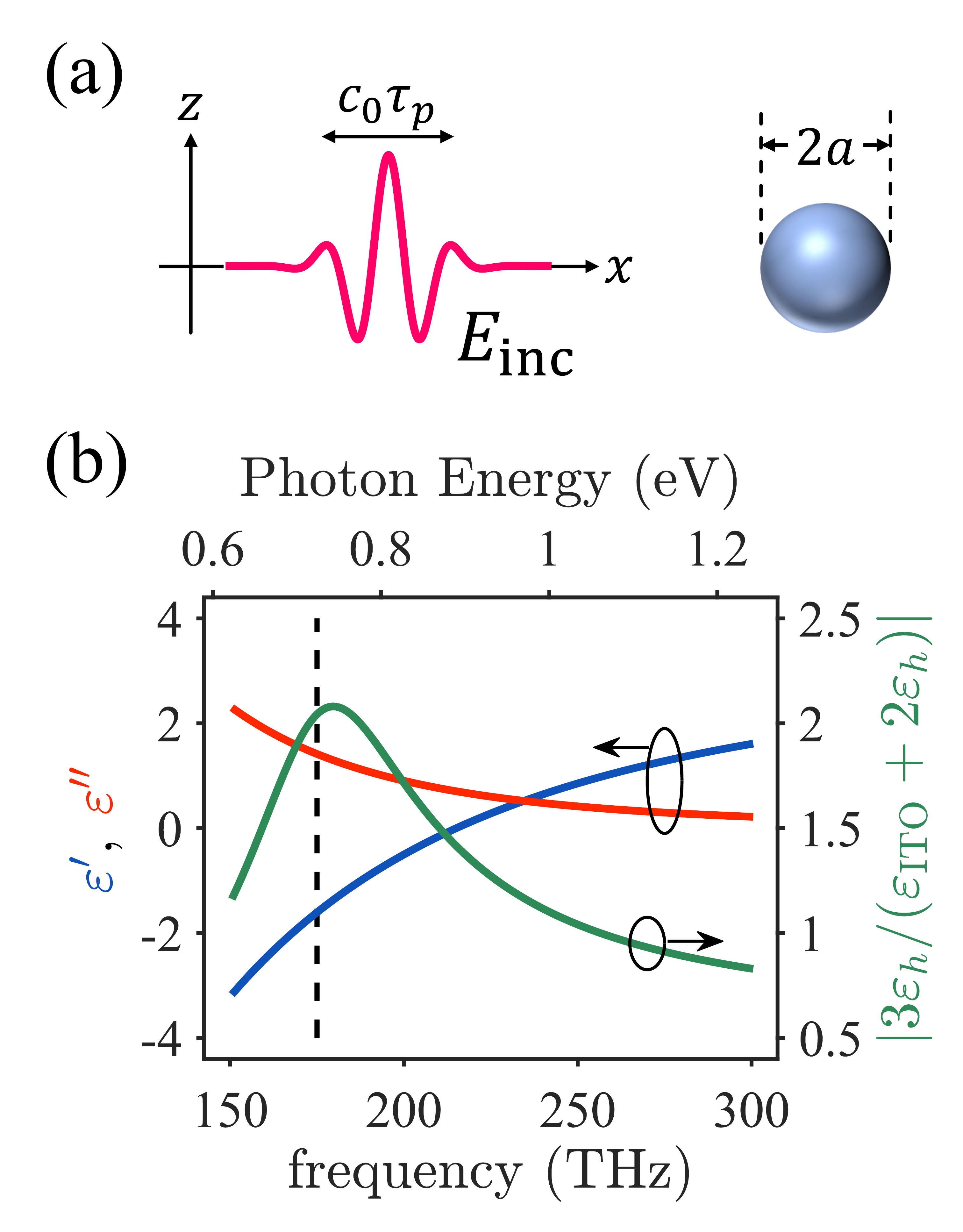}
\caption{\XYZ{}{(a) Schematics of the configuration considered here. (b) The real (blue) and imaginary (red) part of the permittivity of the ITO NP, and the absolute value of the prefactor of the linear quasi-static solution for the local field inside a subwavelength ITO sphere (green) at $T_e = 300$ K. The vertical dashed line represents the central frequency ($175$ THz) used in the simulations.}}
\label{fig:sch_eps_Enp_Einc_frac}
\end{figure}

Now that we have obtained the dynamics of the DM elements, to obtain the electron and electromagnetic field dynamics, one generally requires a self-consistent numerical iteration scheme involving the Boltzmann equation, Maxwell's equations, and Eq.~\eqref{eq:drhoEE_dt}. For simple geometries, it is possible to circumvent the direct solution of Maxwell's equations by leveraging existing known solutions. As an illustrative example, we consider a single small spherical NP with a radius $a$ much smaller than the central wavelength of the incident pulse $\lambda_0$, \XYZ{}{embedded in a homogeneous host medium of permittivity $\varepsilon_h$, see Fig.~\ref{fig:sch_eps_Enp_Einc_frac}(a)}. This enables employing the quasi-electrostatic approximation of Maxwell's equations~\cite{Jackson-book}, i.e., $\partial {\bf B}/\partial t \approx 0$. 
\XYZ{}{We decompose the total polarization density into non-dispersive and dispersive contributions. The non-dispersive part accounts for the response of the bound electrons and is assumed, due to the interest in near infrared frequencies, to remain unchanged under illumination; it is characterized by a constant susceptibility $\varepsilon_\infty - 1$, where $\varepsilon_\infty$ is the permittivity at infinite frequency. The dispersive contribution, denoted by ${\bf P}_\textrm{NP}$, arises from the conduction electrons and is obtained from Eq.~\eqref{eq:Pnp_Esp}.} The advantage of this configuration is that the local electric field ${\bf E}_\textrm{NP}$, \XYZ{}{(the dispersive part of)} the polarization density ${\bf P}_\textrm{NP}$ as well as \XYZ{}{the electric displacement ${\bf D}_\textrm{NP} = (\varepsilon_0 \varepsilon_\infty {\bf E}_\textrm{NP} + {\bf P}_\textrm{NP})$} inside the NP are uniform and that they relate to the incident electric field via~\cite{Jackson-book} 
\begin{align}\label{eq:E_NP}
{\bf E}_\textrm{NP}(t) = \dfrac{1}{\varepsilon_\infty + 2 \varepsilon_h}\left[3 \varepsilon_h {\bf E}_\textrm{inc}(t) - \dfrac{{\bf P}_\textrm{NP}(t)}{\varepsilon_0} \right],
\end{align}
\XYZ{}{where $\varepsilon_0$ is the vacuum permittivity. In the weak-field and continuous-wave limit, the Fourier transform of ${\bf P}_\textrm{NP}$ (Eq.~\eqref{eq:Pnp_Esp}) recovers the Drude susceptibility multiplied by the Fourier-transformed local electric field. The resulting permittivity can be written as~\footnote{In Eq.~\eqref{eq:H_int}, the local electric field is assumed in $z$-direction. In this regard, Eq.~\eqref{eq:eps_WFCW} represents the $zz$-component of the permittivity tensor.} (see also details in Ref.~\cite{Sarkar-Un-Sivan-DM})
\begin{align}\label{eq:eps_WFCW}
\varepsilon(\omega,T_0) = \varepsilon_\infty + \dfrac{V}{2\varepsilon_0\hbar}\int d\mathcal{E}d\mathcal{E}' \dfrac{ \rho_e(\mathcal{E}) \rho_e(\mathcal{E}') (\mathcal{E} - \mathcal{E}') |D(k(\mathcal{E}),k(\mathcal{E}'),L)|^2 (f(\mathcal{E},T_0) - f(\mathcal{E}',T_0) ) }
{ \omega^2 - (\mathcal{E} - \mathcal{E}')^2/\hbar^2 - 2i\eta_{\mathcal{E}\mathcal{E}'}\omega}.
\end{align}
Here, $f(\mathcal{E},T_0)$ is the electron distribution at $T_0 = 300$ K. The ITO permittivity (shown in Fig.~\ref{fig:sch_eps_Enp_Einc_frac}(b)) incorporates the finite-size effect through the energy-space dipole matrix element $|D|$ (Eq.~\eqref{eq:d2D}), leading to a larger imaginary part compared to the bulk. Consistently, the Fourier transform of Eq.~\eqref{eq:E_NP} reproduces the standard quasi-electrostatic solution for the internal field of a small spherical NP~\cite{Bohren-Huffman-book}, exhibiting a relatively broad resonance centred at $\sim180$ THz, see Fig.~\ref{fig:sch_eps_Enp_Einc_frac}(b).} 

The electric field of the incident pulse is assumed to have a Gaussian envelope, i.e., ${\bf E}_\textrm{inc}(t) = \hat{{\bf z}} E_0 e^{-2 \ln 2 (t/\tau_p)^2}\cos(\omega_0 t)$, where $E_0$ is the peak amplitude of the incident pulse. \XYZ{We set $\tau_p = 5$ femtoseconds to be the pulse duration, and $\sim 0.723$ eV or equivalently, $\omega_0/2\pi = 175$ THz (a $\sim 6$ femtoseconds long cycle) as the central frequency.}{We set the pulse duration to $\tau_p = 5$ fs and choose a central frequency $\omega_0/2\pi = 175$ THz (equivalently $\sim 0.723$ eV), close to the resonance (see Fig.~\ref{fig:sch_eps_Enp_Einc_frac}(b)), corresponding to an optical cycle of approximately 6 femtoseconds.} The incident pulse is assumed to be polarized in the $z$ direction without loss of generality. However, for simplicity, from now on we suppress the vectorial nature of the electric fields.

\section{Results}\label{sec:Results}

\subsection{Term comparison}
\begin{figure}[h]
\centering
\includegraphics[width=\textwidth]{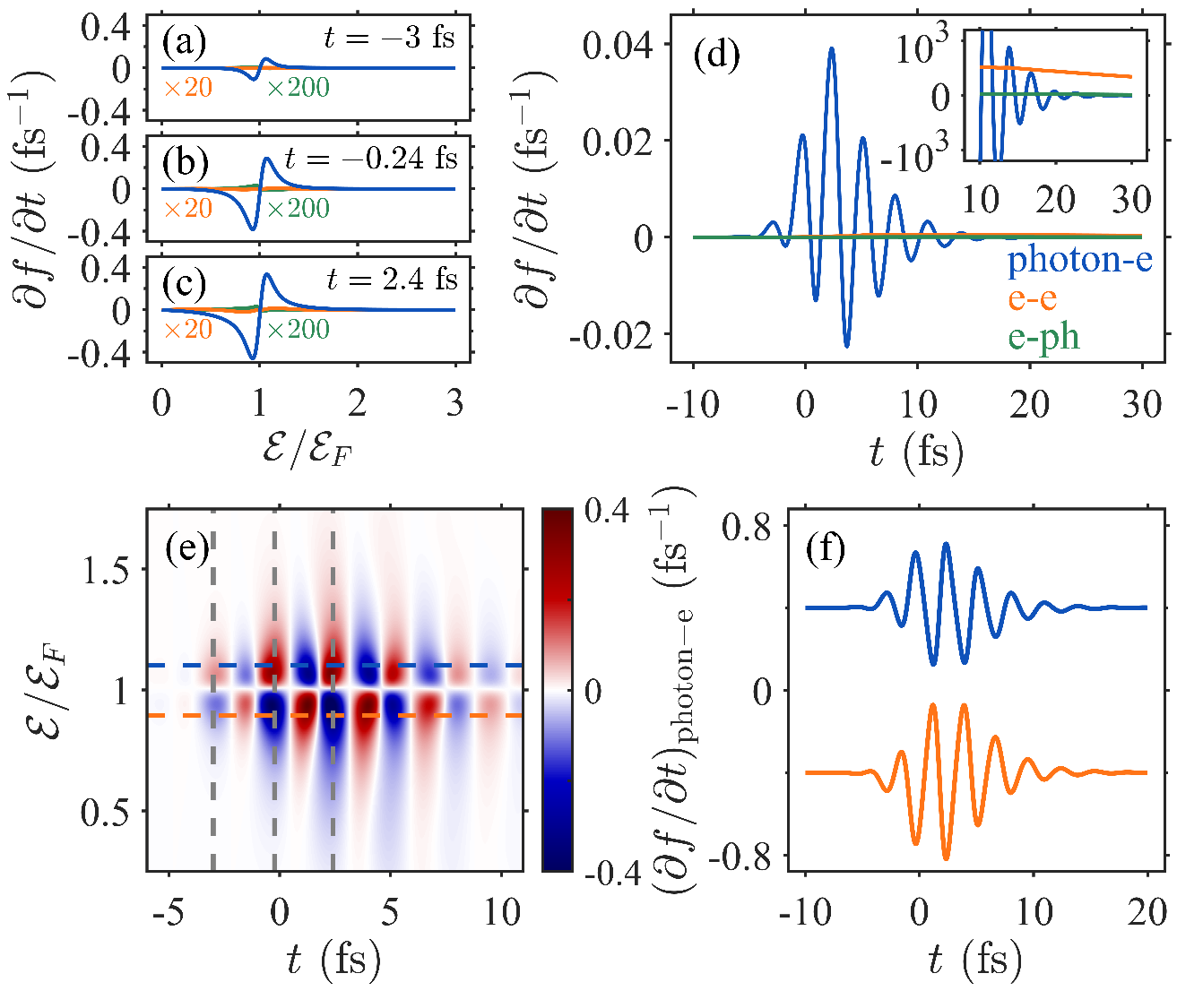}
\caption{(a)-(c) The $\epht$ (blue), $\ee$ (orange), and $\eph$ (green) terms for \XYZ{$E_0 = 0.6\times 10^9$ V/m}{$E_0 = 0.6$ V/nm} at $t = -3.2$ fs, $-0.5$ fs, and $2.2$ fs. \XYZ{}{(d) The time-dependence of photon-e (blue), e-e (orange), and e-ph (green) terms at $\mathcal{E} = \mathcal{E}_F + \hbar\omega_0/2$. The insets show the zoom-in for $t>10$ fs.} (e) The color map of the photon-e term (Eq.~\eqref{eq:dfdt_exc_E}). The black dashed lines label the time sections of (a)-(c). The blue and orange dashed lines label the energy sections shown in (f). (f) The electron distribution dynamics at $\e = \e_F \pm \hbar \omega_0/8$ (blue and orange).
}\label{fig:dfdt_0.6Vm-1}
\end{figure}

We first plot in Fig.~\ref{fig:dfdt_0.6Vm-1}(a)-(c) the photon-e, $\ee$, and $\eph$ terms of Eq.~\eqref{eq:relax_op_aa} for \XYZ{$E_0 = 0.6 \times 10^9$ V/m}{$E_0 = 0.6$ V/nm}. One can see that, due to the dominance of near-diagonal terms in the dipole transition elements (see Fig.~\ref{fig:mat_ele_eng}), the photon-e term exhibits distinct thermal-like features~\cite{Dubi-Sivan,Govorov_ACS_phot_2017} in the vicinity of the Fermi energy (see Fig.~\ref{fig:dfdt_0.6Vm-1}(a)-(c)). Additionally, the effect of electron collisions with other electrons is an order of magnitude weaker, and the effect of electron collisions with phonons is a further order of magnitude smaller. \XYZ{}{Fig.~\ref{fig:dfdt_0.6Vm-1}(d) further shows that throughout most of the pulse duration, the photon–e term overwhelmingly dominates over both scattering processes.} Consequently, the ultrafast (few-femtosecond) evolution of the electron distribution is primarily driven by the photon–electron interaction. \XYZ{}{The e–e term becomes appreciable only toward the tail of the pulse ($t \gtrsim 15$ fs), after the peak of the optical excitation (see inset of Fig.~\ref{fig:dfdt_0.6Vm-1}(d)).} In~\cite{single_cycle_nlty_Letter}, we focused on the analysis of the overall dynamics and the two main competing contributions to this dominant photon-e term, namely, absorption and stimulated emission. Here, instead, we focus on aspects not discussed in~\cite{single_cycle_nlty_Letter}; specifically, after justifying the neglect of the contribution of spontaneous emission to photon-e interactions (see Appendix.~\ref{app:PL}), we here dwell on the microscopic details of the (stimulated) photon-e interactions, namely, their energy and time dependence. This will include the relation of this term to the coherence dynamics and the origins of the super-linear scaling of the photon-e term with the illumination intensity.

\begin{figure}[h]
\centering
\includegraphics[width=\textwidth]{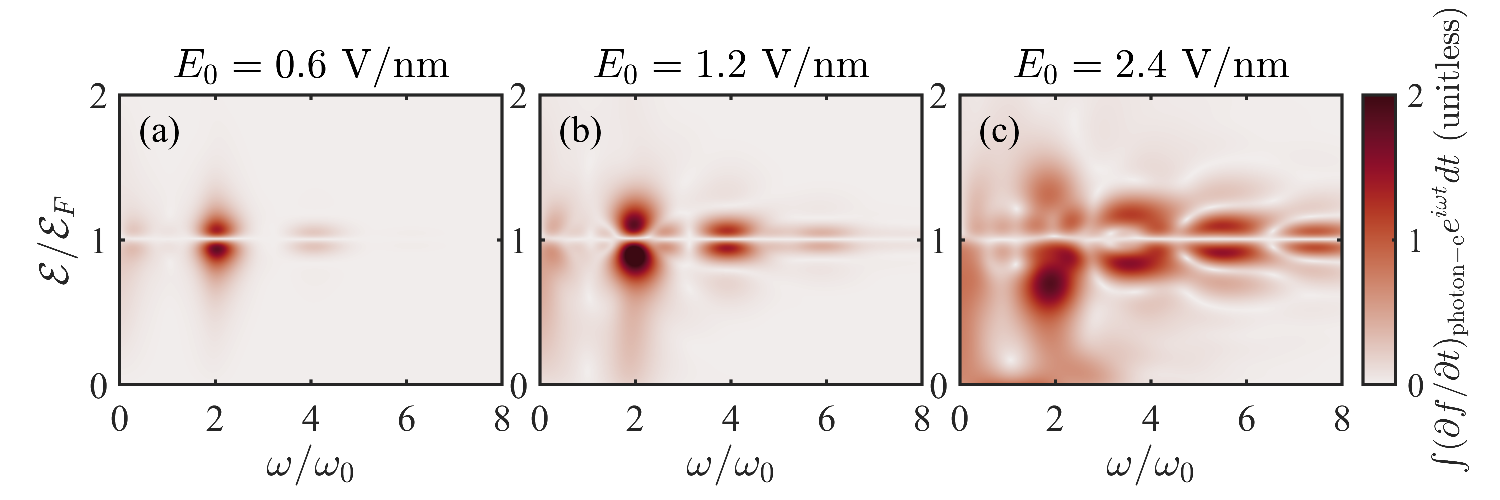}
\caption{(a)-(c) The Fourier transform of the photon-e term for $E_0 = 0.6$ V/nm, $1.2$ V/nm, and $2.4$ V/nm. 
}\label{fig:dfdt_exc_FT_REF}
\end{figure}

\begin{figure}[h]
\centering
\includegraphics[width=0.7\textwidth]{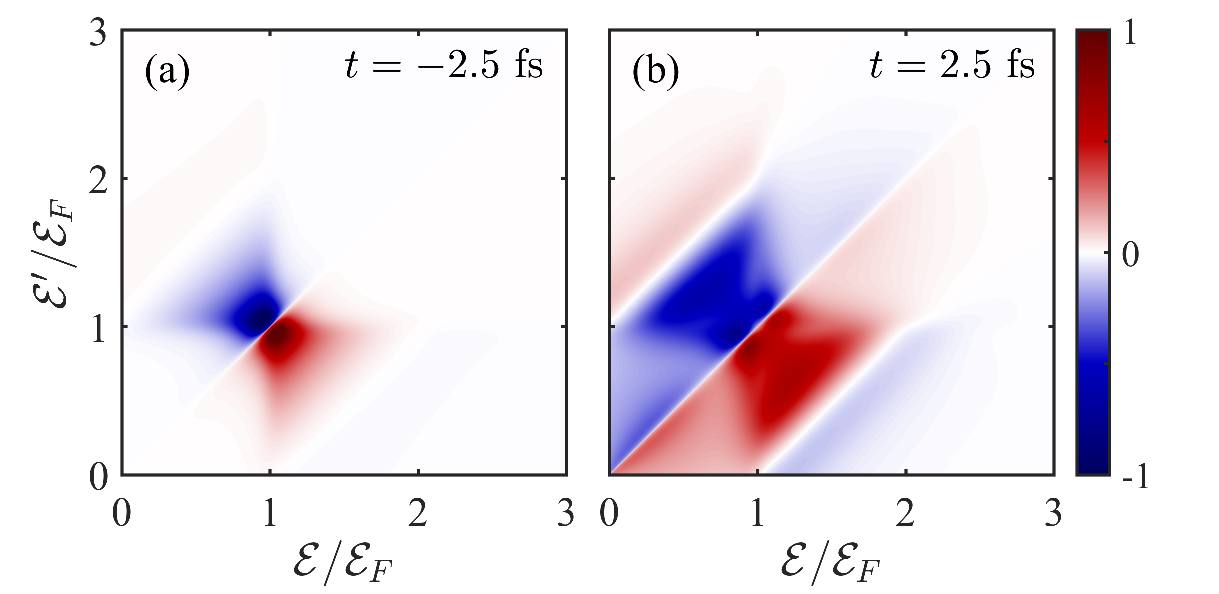}
\caption{The imaginary part of the density matrix for \XYZ{$E_0 = 1.2 \times 10^9$ V/m}{$E_0 = 1.2$ V/nm} at (a) $t = -2.5$ fs and (b) $2.5$ fs, respectively. }
\label{fig:antisym_rho_1.2Vm-1_REF}
\end{figure}

Thus, specifically, Figs.~\ref{fig:dfdt_0.6Vm-1}\XYZ{(d) and (e)}{(e) and (f)} reveal time-oscillations at frequency $\sim 2 \omega_0$ (i.e., with a period of $\sim 3$ fs) in this term. These arise from the product between the $\pm \omega_0$ components of the real electric field and the DM off-diagonal elements, which themselves oscillate at $\pm \omega_0$ as driven by the field (see Eqs.~\eqref{eq:dfdt_exc_E} and~\eqref{eq:drhoEE_dt}). This $\sim 2 \omega_0$ oscillation in the photon-e term, in turn, induces similar oscillations in the distribution. Simultaneously, cross-terms between the $\pm \omega$ components of $E(t)$ and the $\mp \omega$ components of the DM off-diagonal elements generate a zero-frequency component in the photon-e term, resulting in a non-vanishing, time-averaged change in the distribution. As the field intensity increases, the electron distribution oscillation stimulates higher-order oscillations in the off-diagonal density matrix at $\sim \pm 3 \omega_0$, $\sim \pm 5 \omega_0$, etc., via Eq.~\eqref{eq:rho_EOM_k}. This cascade gives rise to additional photon-e and distribution oscillations at $4 \omega_0$, $6\omega_0$, and beyond, see Fig.~\ref{fig:dfdt_exc_FT_REF}. These high-frequency oscillations are typically neglected in the rotating wave approximation; however, in the current context, they signify the crucial role played by the off-diagonal elements of the density matrix in the time evolution. Notably, the off-diagonal elements are mostly dominated by those near the diagonal, with magnitudes comparable to the diagonal elements. More specifically, only off-diagonal elements near the Fermi energy are non-zero at the leading edge of the pulse, while elements further below the Fermi energy become pronounced towards the trailing edge of the pulse, see Fig.~\ref{fig:antisym_rho_1.2Vm-1_REF}.

\begin{figure}[h]
\centering
\includegraphics[width=0.6\linewidth]{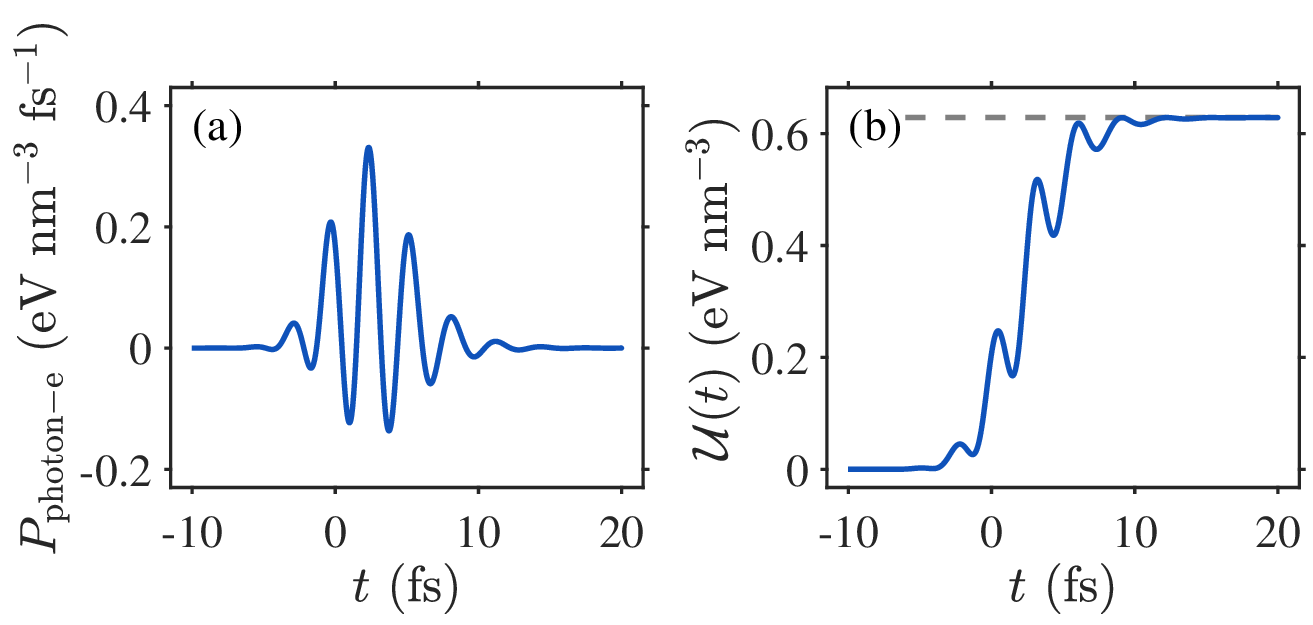}
\caption{Temporal evolution of (a) the net energy (density) absorption rate $P_\epht$~(\ref{eq:P_photon-e}), 
and (b) the cumulative absorbed energy density $\mathcal{U}$~\eqref{eq:U_t}, for a peak field strength of $E_0 = 1.2$ V/nm. 
}\label{fig:P_phton_e_csum}
\end{figure}

Next, we integrate the photon-e term (Eq.~\eqref{eq:dfdt_exc_E}) over energy, resulting with the instantaneous net absorption (density) rate of photon energy by the electron subsystem, defined as 
\begin{align}\label{eq:P_photon-e}
P_\textrm{photon-e}(t) = \displaystyle\int_0^{\e_\textrm{max}} d\e \rho_e(\e) \e\left(\dfrac{\partial f}{\partial t}\right)_\epht.
\end{align}
Fig.~\ref{fig:P_phton_e_csum}(a) reveals time-oscillations of $P_\textrm{photon-e}$ at approximately twice the driving frequency ($\sim 2 \omega_0/(2\pi)\sim 3$ fs), as seen in Figs.~\ref{fig:dfdt_0.6Vm-1}(d) and (e). \XYZ{}{It can be shown that the expression~\eqref{eq:P_photon-e} agrees with the power density of the field doing work on electrons, expressed as $E_\textrm{NP}(t) \cdot {\partial P_\textrm{NP}(t)}/{\partial t}$. This agreement confirms that energy conservation is intrinsically satisfied in our model, without requiring any transient adjustment of the normalization as employed in some previous studies~\cite{delFatti_nonequilib_2000,Un-Sarkar-Sivan-LEDD-II}. Furthermore, this agreement shows that the expression is of a general validity, i.e., both for the linear and nonlinear regimes, and motivated extending it to classical \XYZ{}{thermal} modelling of $E_\textrm{NP}(t) \cdot {\partial P_\textrm{NP}(t)}/{\partial t}$ in Ref.~\cite{single_cycle_nlty_Letter}. In that respect, this quantity correspondence to the ``fast nonlinearity'' discussed in~\cite{Khurgin_Kinsey_LPR} which represents ``acquired'' (but not truly absorbed) energy. } 

Notably, although $P_\textrm{photon-e}(t)$ becomes repeatedly negative, its average is positive, so the cumulative energy transferred to the electron subsystem as manifested via the time integral  
\begin{align}\label{eq:U_t}
\mathcal{U}(t) = \int_{-\infty}^t P_\textrm{photon-e}(t') dt' 
\end{align}
exhibits a net increase over time, see Fig.~\ref{fig:P_phton_e_csum}(b). Interestingly, the behaviour of this time integral $\mathcal{U}(t)$ resembles the permittivity dynamics (see~\cite[Fig.~4(a)-(b)]{single_cycle_nlty_Letter}); this further emphasizes the absorptive nature of the nonlinearity of ITO. Future work would be aimed to establishing a formal link between these quantities. 
\XYZ{}{More generally, this quantity corresponds to the ``slow nonlinearity'' discussed in~\cite{Khurgin_Kinsey_LPR}, which is seen here to be decorated by oscillations originating from the ``fast nonlinearity''. }

\XYZ{}{As a final comment associated with the absorption, we note that our calculations imply that a few eV are absorbed per $\textrm{nm}^3$. Thus, for the $4$ nm sphere studied here 
and for an electron density of $1 / \textrm{nm}^3$, the sphere has 33 electrons. Thus, each of them absorbs 0.03 to 0.15 eV; this prediction is not sensitive to the illuminating pulse duration. This result is not to be confused with the approximation in~\cite{Khurgin_Kinsey_high_nlty} of 1-4 eV acquired per electron via the ``fast nonlinearity'' occurring during the illumination stage.}

\XYZ{It can be shown that the expression~\eqref{eq:P_photon-e} agrees with the power density of the field doing work on electrons, expressed as $E_\textrm{NP}(t) \cdot {\partial P_\textrm{NP}(t)}/{\partial t}$. This agreement confirms that energy conservation is intrinsically satisfied in our model, without requiring any transient adjustment of the normalization as employed in some previous studies}{}\XYZ{(e.g.~\cite{Un-Sarkar-Sivan-LEDD-II} and~\cite{delFatti_nonequilib_2000})}{}\XYZ{. Furthermore, this agreement shows that the expression is of a general validity, i.e., both for the linear and nonlinear regimes, and, in fact, it also extends to classical modelling of $E_\textrm{NP}(t) \cdot {\partial P_\textrm{NP}(t)}/{\partial t}$. Moreover, this expression is the source term used in the thermal model described in Ref.}{}\XYZ{~\cite{single_cycle_nlty_Letter}}{}.


\subsection{Excited state absorption}\label{sec:ex_state_abs}
\begin{figure}[t]
\centering
\includegraphics[width=0.6\linewidth]{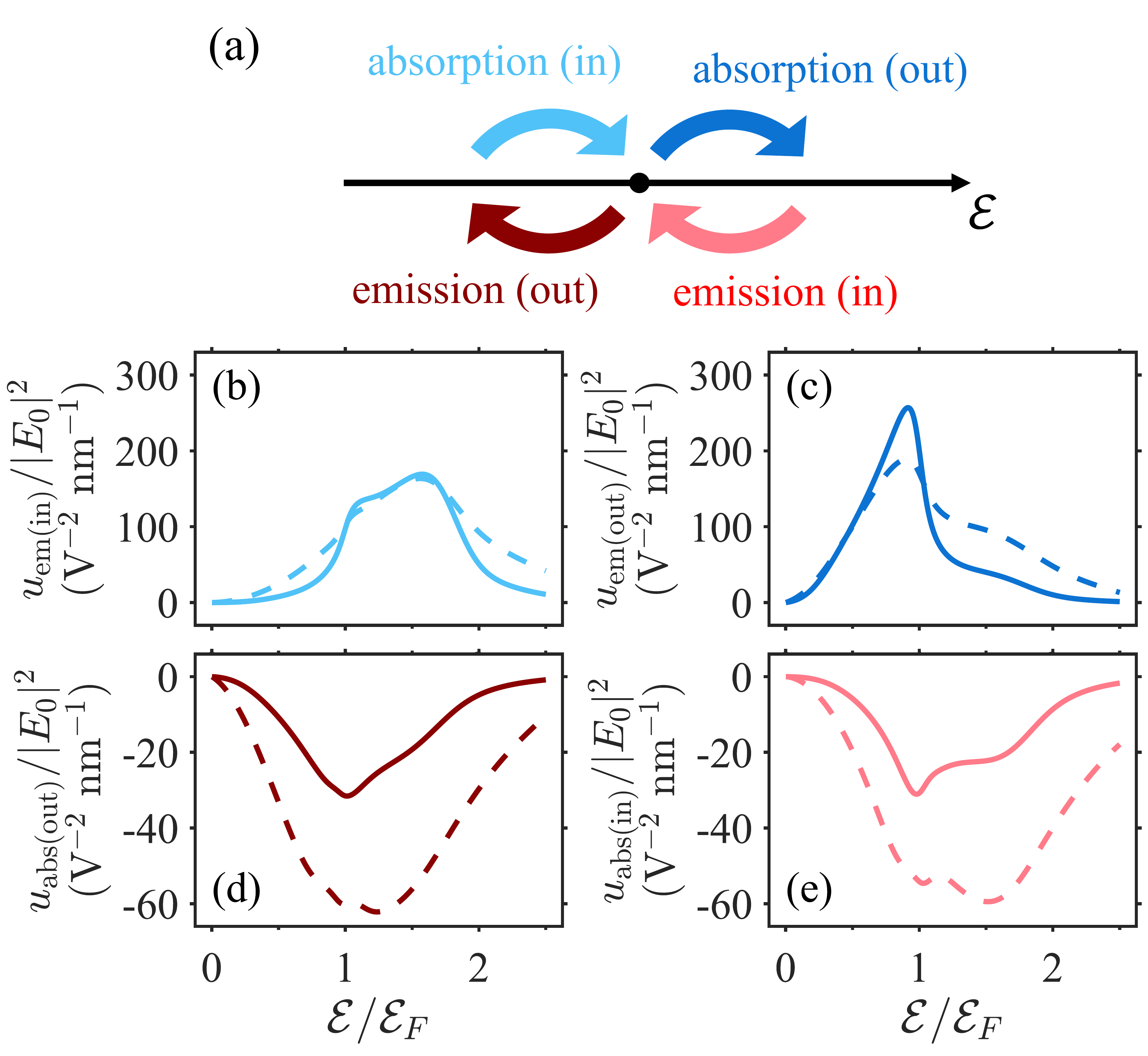}
\caption{(a) Schematic of the decomposition of the photon-e interaction term~(\ref{eq:dfdt_exc_E}) into four distinct contributions. \XYZ{(b) $u_\text{abs(out)}$, and (c) $u_\text{em(out)}$}{The integrands (b) $u_\text{abs(in)}$, (c) $u_\text{abs(out)}$, (d) $u_\text{em(out)}$, and (e) $u_\text{em(in)}$}, (normalized by $|E_0|^2$) correspond to the energy-resolved, time-integrated contributions from each process for $E_0 = 0.6$ V/nm (solid lines) and $1.2$ V/nm (dashed lines).
}
\label{fig:Du_abs_emi_out}
\end{figure}

To gain deeper insight into the scaling of (stimulated) absorption with incident power, and specifically, its superlinear scaling with $|E_0|^2$ observed in Fig.~4(e) of Ref.~\cite{single_cycle_nlty_Letter}, we decompose the $\epht$ interaction term~(\ref{eq:dfdt_exc_E}) for a specific energy state $\e$ into four distinct contributions: 
exciting electrons from lower-energy states via absorption (denoted as ``absorption (in)''); exciting electrons to higher-energy states via absorption (denoted as ``absorption (out)''); de-exciting electrons from higher-energy states via emission (denoted as ``emission (in)''); de-exciting electrons to lower-energy states via emission (denoted as ``emission (out)''), see Fig.~\ref{fig:Du_abs_emi_out}(a). This decomposition can be performed by rewriting the source term in Eq.~\eqref{eq:drhoEE_dt} according to
\begin{multline}\label{eq:decomp_f}
f(\e,t)-f(\e',t) = f(\e,t)(1 - f(\e',t))\Theta(\e'-\e) + f(\e,t)(1 - f(\e',t))\Theta(\e - \e')
\\
+ (1 - f(\e,t))f(\e',t)\Theta(\e - \e') + (1 - f(\e,t))f(\e',t)\Theta(\e'-\e),
\end{multline}
\XYZ{}{where $\Theta$ is the Heaviside step function.} If we then multiply each of these four terms by the electron energy and the eDOS, and then integrate them over time (rather than on electron energy as in Eq.~(\ref{eq:P_photon-e})), the resulting quantities represent the individual contributions of each process to the total energy 
change of the electron system, i.e., 
\begin{align}
u_{\substack{
\textrm{abs (out)}\\
\textrm{em (out)}\\
\textrm{abs (in)}\\
\textrm{em (in)}}} 
= -\dfrac{i \pi V}{8\hbar} \int_{-\infty}^{\infty}
\int_0^{\e_\textrm{max}} 
E_\textrm{NP}(t)(\e-\e')\rho_e(\e)\rho_e(\e')\left(\overline{d_{\e'\e} \rho_{\e\e'}} - \overline{d_{\e\e'} \rho_{\e'\e}}\right)_{\substack{
\textrm{abs (out)}\\
\textrm{em (out)}\\
\textrm{abs (in)}\\
\textrm{em (in)}}} d\e' dt. 
\end{align}
so that 
\begin{align}
\mathcal{U}_{\substack{
\textrm{abs (out)}\\
\textrm{em (out)}\\
\textrm{abs (in)}\\
\textrm{em (in)}}} = \int_0^{\e_\textrm{max}}   u_{\substack{
\textrm{abs (out)}\\
\textrm{em (out)}\\
\textrm{abs (in)}\\
\textrm{em (in)}}} (\e) d\e,  
\end{align}
i.e., the sum of all four contributions, $u_\textrm{abs,in}$, $u_\textrm{abs,out}$, $u_\textrm{em,in}$, and $u_\textrm{em,out}$, integrated over $\mathcal{E}$ recovers the total energy transferred via the photon–e interaction, i.e., Eq.~\eqref{eq:U_t}. \XYZ{Figs.~\ref{fig:Du_abs_emi_out}(b) and (c) show the contributions from absorption (out) and emission (out) for $E_0 = 0.6$ V/nm and $1.2$ V/nm.}{Figs.~\ref{fig:Du_abs_emi_out}(b)-(e) show the contributions from absorption (in and out) and emission (in and out) for $E_0 = 0.6$ V/nm and $1.2$ V/nm.} At the lower illumination level ($E_0 = 0.6$ V/nm), absorption is dominated by excitation from below the Fermi level to higher-energy states. As the field strength increases, this contribution weakens, and excitations from already excited electrons ($\mathcal{E} > \mathcal{E}_F$) to even higher levels become more significant\XYZ{, see Fig.~\ref{fig:Du_abs_emi_out}(b)}{}. 
The corresponding absorption (in) process exhibits behavior similar to that of the absorption (out) counterparts. In particular, the energy-resolved contribution $u_\textrm{abs (in)}(\mathcal{E})$ peaks near $\mathcal{E}_F +\hbar\omega_0$. As the illumination level increases, the peak slightly decreases, while the contributions at higher energies ($\mathcal{E} \gtrsim \mathcal{E}_F + \hbar\omega_0$) become more prominent, again indicating the onset of excited state absorption processes\XYZ{}{, see Figs.~\ref{fig:Du_abs_emi_out}(b) and (c)}. 
This behavior indicates that the superlinear scaling of (stimulated) absorption with incident power originates from 
a non-simultaneous two-photon absorption process, analogous to excited-state absorption in atomic systems. It reflects the increasing availability of high-energy absorption channels at elevated field strengths and contrasts sharply with the saturation behavior typical of two-level systems, where absorption eventually becomes sublinear due to ground-state depletion. 

In contrast, the emission (out) (and (in)\XYZ{, not shown}{}) contributions, when normalized by $|E_0|^2$, increase with incident power, but their energy dependence remains largely unchanged, see \XYZ{Fig.~\ref{fig:Du_abs_emi_out}(c)}{Figs.~\ref{fig:Du_abs_emi_out}(d) and (e)}. This is consistent with the observed $|E_0|^4$ scaling of emission shown in Fig.~4(e) of Ref.~\cite{single_cycle_nlty_Letter}. Notably, such behavior can only be captured by explicitly tracking the time-dependent population dynamics and cannot be reproduced by phenomenological models.



\XYZ{As a final comment associated with the absorption, we note that our calculations imply that a few eV are absorbed per $\textrm{nm}^3$. Thus, for the $4$ nm sphere studied here 
and for an electron density of $1 / \textrm{nm}^3$, the sphere has 33 electrons. Thus, each of them absorbs 0.03 to 0.15 eV; this prediction is not sensitive to the illuminating pulse duration. This result is very different from the approximation in~\cite{Khurgin_Kinsey_high_nlty} of 1-4 eV absorbed per electron.}{}

\subsection{Local field}
\begin{figure}[h]
\centering
\includegraphics[width=1\linewidth]{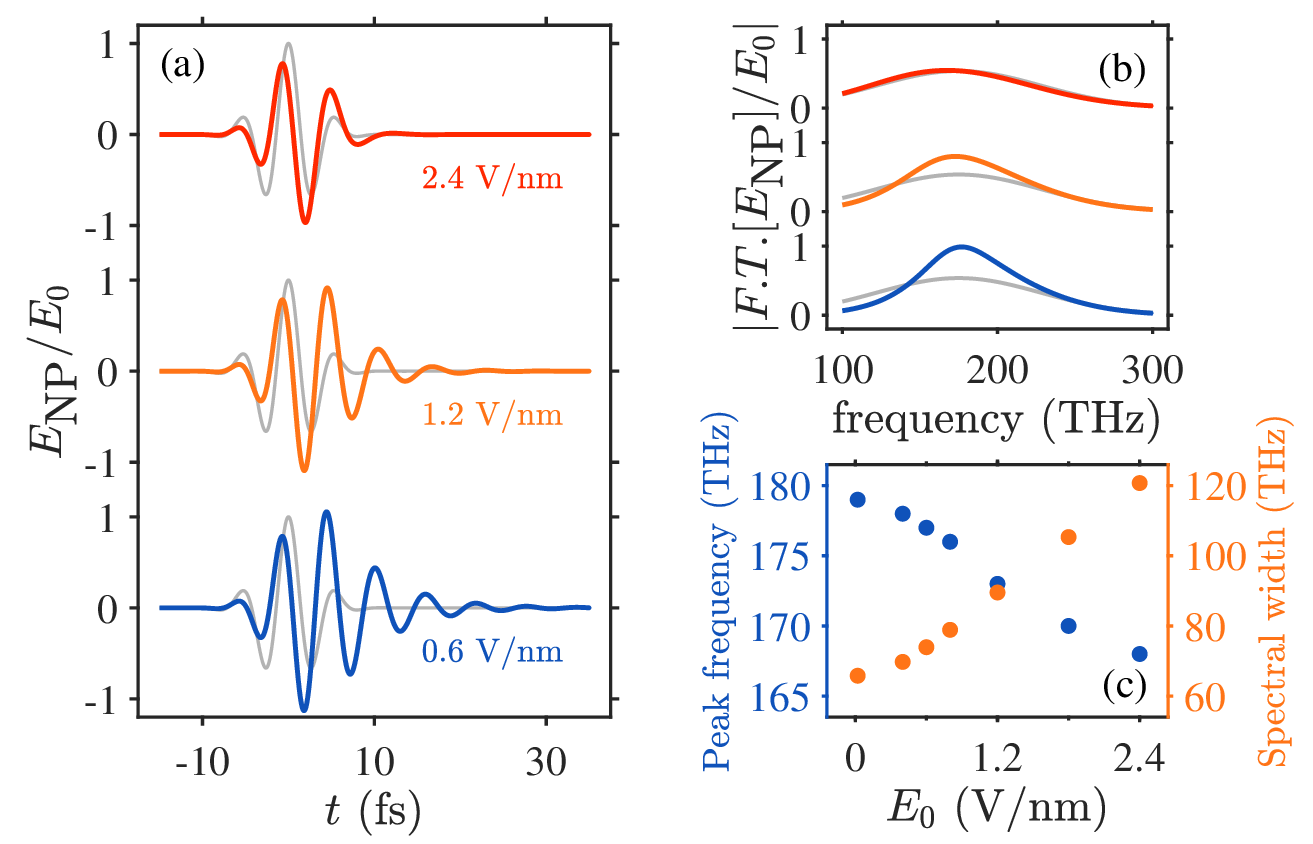}
\caption{(a) Local electric field for $E_0 = 0.6$ V/nm (blue), $1.2$ V/nm (orange), and $2.4$ V/nm (red) vs. the corresponding incident fields (grey). (b) The corresponding spectra (i.e., the Fourier transform, denoted by $F.T.$) for the local and incident electric field in (a). (c) The peak frequency (blue dots) and spectral width (orange dots) of the local electric fields as a function of the incident field strength. 
}
\label{fig:local_field}
\end{figure}

Finally, we focus on the temporal variation of the local electric field (Eq.~\eqref{eq:E_NP}), see Fig.~\ref{fig:local_field}(a). Due to the resonance, the local field exhibits a longer temporal duration compared to the incident field. This extension gradually diminishes with increasing incident intensity, and is accompanied by spectral broadening (self-phase modulation, Fig.~\ref{fig:local_field}(b)) and a red-shift (Fig.~\ref{fig:local_field}(c)). Such behavior is consistent with the expected increase in the real part of the ITO permittivity under high-field excitation (see, e.g.,~\cite{Kinsey-Nat-Rev-Mater-2019,Zhou_freq_trans_ENZ_2020,Pang-AFC-NL-2021,Un-Sarkar-Sivan-LEDD-II}).

\begin{figure}[h]
\centering
\includegraphics[width=1\linewidth]{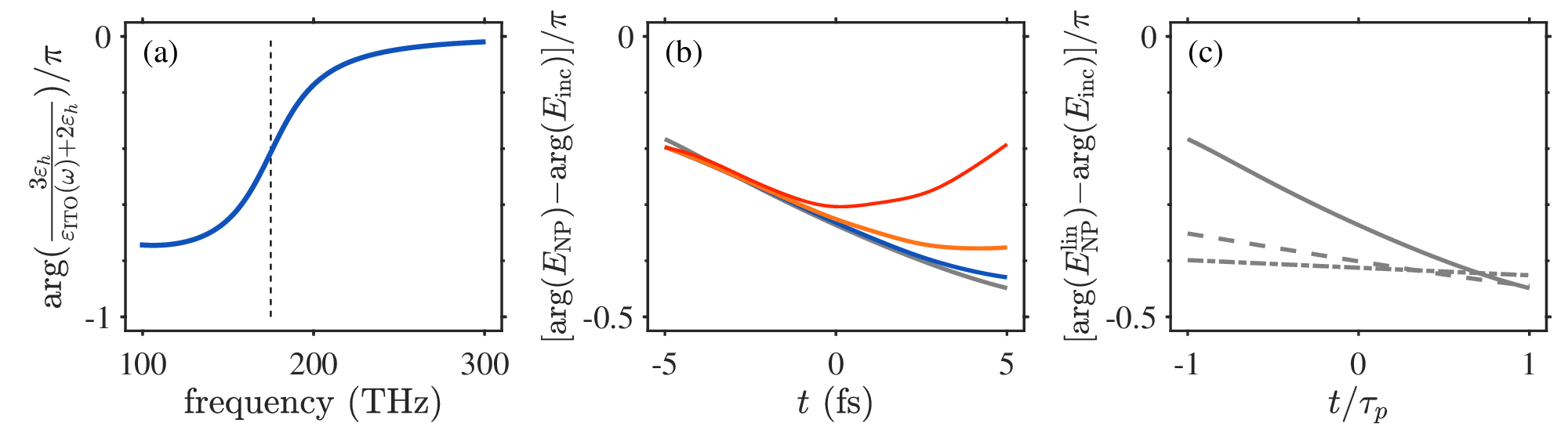}
\caption{(a) The phase of the prefactor of the linear quasi-static solution for the local field inside a subwavelength dielectric sphere; the vertical dashed line represents the central frequency ($175$ THz) used in the simulations. (b) The phase difference between the local field $E_\textrm{NP}(t)$~\eqref{eq:E_NP} and the incident field for the three cases shown in Fig.~\ref{fig:local_field}. \XYZ{(c) Same as (b), but showing the electric field under weak illumination (i.e., in the linear regime) for incident pulses with durations of $5$ fs (solid), $20$ fs (dashed), and $80$ fs (dash-dotted)}{(c) The phase difference between the local and incident fields under weak illumination (i.e., in the linear regime) for three pulse durations: 5 fs (solid), 20 fs (dashed), and 80 fs (dash-dotted)}. 
}
\label{fig:Enp_Einc_ph}
\end{figure}

In Fig.~\ref{fig:Enp_Einc_ph}, we show the corresponding phase difference between these quantities (the local field $E_\text{NP}$ and the incident field $E_\text{inc}$) obtained using the Hilbert transform~\cite{hilbert_transf_bracewell}. \XYZ{From the quasi-static expression for the (linear) response of a deep subwavelength dielectric sphere $3 \varepsilon_h / (\varepsilon_\textrm{ITO}(\omega_0) + 2 \varepsilon_h)$, 
one expects a phase of approximately $- 0.4 \pi$,}{ At the central frequency, where $\varepsilon_\mathrm{ITO}\approx -1.6+1.4i$ (see Fig.~\ref{fig:sch_eps_Enp_Einc_frac}(b)), the quasi-static linear response of a deeply subwavelength sphere predicts a phase shift of $\arg(3 \varepsilon_h / (\varepsilon_\textrm{ITO}(\omega) + 2 \varepsilon_h))\approx -0.4\pi$,} as marked by the dashed vertical line shown in Fig.~\ref{fig:Enp_Einc_ph}(a). However, the actual phase shifts extracted from Fig.~\ref{fig:local_field}(a), vary between $\sim -0.2 \pi$ and $-0.4 \pi$, as shown in Fig.~\ref{fig:Enp_Einc_ph}(b). This discrepancy stems from the broadband nature of the incident $5$ fs pulse. To illustrate this, Fig.~\ref{fig:Enp_Einc_ph}(c) plots the phase difference between $E_\textrm{NP}$ and $E_\textrm{inc}$ in the linear regime for growing pulse durations. This shows that the time dependence of the phase difference becomes weaker and gradually converges toward the expected quasi-static value as the pulse duration increases. This confirms that the observed discrepancy is indeed a consequence of the pulse's large spectral width.

\begin{figure}[h]
\centering
\includegraphics[width=1\linewidth]{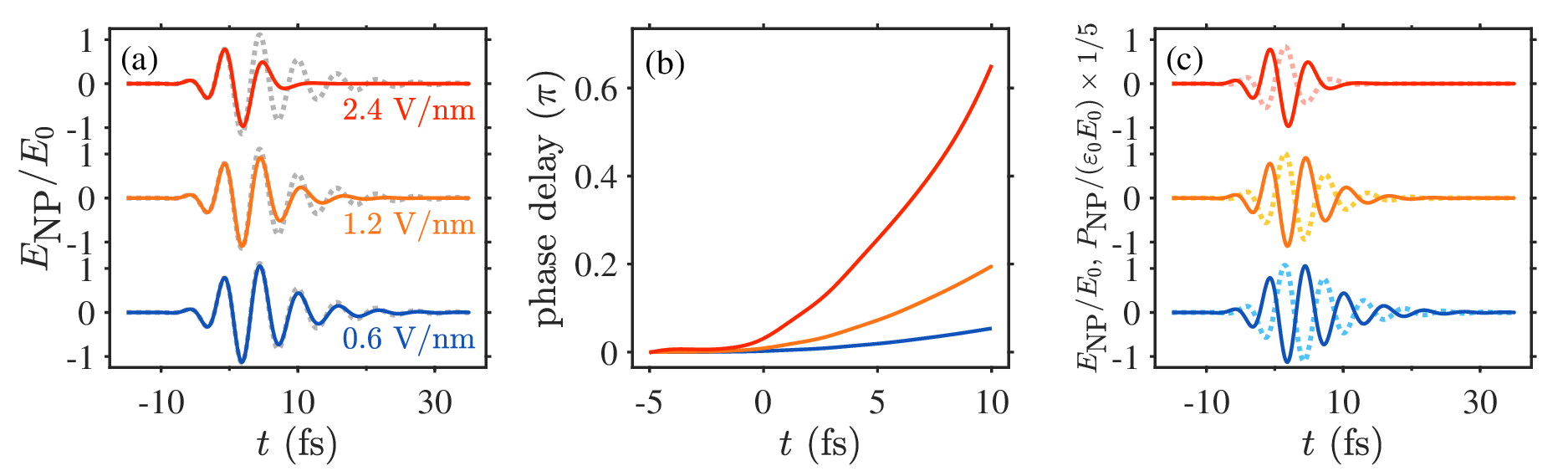}
\caption{(a) Local field (solid lines) vs. the corresponding reference linear field (dotted lines) for the three cases shown in Fig.~\ref{fig:local_field}. (b) The corresponding phase difference between each curve pair in (a). (c) The same as (a) but comparing with the corresponding (scaled) $P_\textrm{NP}$ (dotted lines). }
\label{fig:Enp_Eref_Pnp}
\end{figure}

In order to eliminate the geometric effect in the quasi-static solution on the phase of the local field and isolate the effect of the nonlinearity, we also compare (the phase of) $E_\textrm{NP}$ with a reference linear field. The latter 
is the time-dependent local field at the weak $E_0$ limit~\footnote{It can be obtained either by performing a time-domain simulation with a sufficiently small value of $E_0$, or equivalently through a Fourier transform method: $\textrm{I.F.T}. \left[ \left(3 \varepsilon_h / (\varepsilon_\textrm{ITO}(\omega) + 2 \varepsilon_h)\right)\right] \textrm{F.T.} \left[E_{inc}](\omega)\right](t)$, where F.T. and I.F.T. denote the Fourier and inverse Fourier transforms, respectively. }. Figs.~\ref{fig:Enp_Eref_Pnp}(a)-(b) show that for the phase difference on the leading edge of the pulse ($t < 0$) aligns well with the linear case for all field strengths, while increasing deviations emerge on the trailing edge as the field intensity grows. This trend indicates that the nonlinearity is predominantly absorptive. Importantly, this is different from what was observed for the Kerr (i.e., an instantaneous, non-absorptive cubic) nonlinearity of SiO$_2$ for which the phases of the corresponding fields differ only around the peak intensity, see~\cite{Schultze_Kerr_SiO2}. A corresponding measurement (or TD-DFT calculation) can determine whether the nonlinearity of TCOs is indeed absorptive, as modelled here, or whether it involves virtual transitions, hence, a Kerr nonlinearity, as claimed in~\cite{Narimanov_2025}. It may also reveal the role of multi-photon absorption, as seen in~\cite{Schultze_Kerr_SiO2,Yabana_PRB_2024}. 

Finally, we also plot the phase difference between $E_\textrm{NP}$ and $P_\textrm{NP}$. One can see that in general, these quantities are out of phase, see Fig.~\ref{fig:Enp_Eref_Pnp}(c). This is a signature of the negativity of the real part of the ITO permittivity.



\section{Outlook}\label{sec:conclusion}
\XYZ{}{In this work, starting from the density-matrix equations formulated in momentum space, we develop an energy-space density-matrix framework for conduction electrons driven by intense single-cycle optical pulses. Applying this approach to an ITO nanosphere, we show that the nonlinear response extends well beyond simple population redistribution: coherences between conduction-band states give rise to pronounced oscillatory dynamics and repeated frequency-doubling features in the polarization. We further identify clear signatures of coherence in the absorption and permittivity dynamics. We also reveal the emergence of strong excited-state absorption which is responsible for the superlinear scaling of the absorbed energy; this behaviour is the opposite to the sublinear behavior typically associated with ground-state depletion in two-level systems, and it mediates the sublinear growth of the net absorption. Finally, we systematically characterize the impact of pulse intensity on the temporal duration, spectral broadening and shift, and phase of the local field, thereby establishing the predominantly absorptive nature of the nonlinear response induced by single-cycle excitation in TCOs.}

This work paves the way to more comprehensive studies of the nonlinear optical response for different particle sizes and shapes (as, e.g., in~\cite{Guo_ITO_nanorod_natphoton})
, different material parameters or TCOs (effective mass, non-parabolicity, electron density, ...) 
different geometries (primarily, of layers, which is the geometry of choice in most experiments to date), for different pump wavelengths (e.g., for a $800$ nm, as in~\cite{Shalaev_Segev_nanophotonics_2023})
, probe wavelength and their duration and chirp. It would also be a basis for comparison with different more computationally accessible approaches (e.g., the hydrodynamic approach, see~\cite{Scalora-ITO-2025}), thus, enabling further studies of complex spatio-temporal dynamics of single cycle pulses in TCOs and interplay of coherent wave mixing with permittivity changes studied therein.

Our work should be a cornerstone for the interpretation of existing and future experimental results on TCOs, in particular, in the determination of the rise time~\cite{Boyd_Nat_Phot_2018,Sapienza_2022,Sapienza_double_slit_2022} 
and explanation of the relaxation dynamics~\cite{Shalaev_Segev_nanophotonics_2023}, and for the comparison to similar studies in dielectrics and semiconductors~\cite{Schultze_Kerr_SiO2}. Specifically, to connect to the literature on the interaction of single cycle pulses with atoms, molecules and semiconductors, one should add to our formulation also interband transitions, i.e., one would need to extend the $\epht$ formulation to include both direct and indirect transitions from a (new) block of the density matrix representing the valence band states to the (existing) block representing the conduction band states. Such a formulation would provide a generalization of the formulation of Rossi and Kuhn~\cite{RevModPhys-Rossi-Kuhn} which would then treat the coherence between all states, and a generalization of the Maxwell-Bloch discrete state formulation (e.g., Refs.~\cite{Boyd-book,Hess-book,Peschel_Maxwell_Bloch}) to a continuum of states. It would allow the identification of the relative magnitude of the competing intraband and interband single and multiple photon coherent and incoherent  nonlinearities.

\acknowledgments
I.W.U. was funded by the Guangdong Natural Science Foundation (Grants No. 2024A1515011457). Y.S. was partially funded by a Lower-Saxony-Israel collaboration grant no. 76251-99-7/20 (ZN 3637) as well as an Israel Science Foundation (ISF) grant (340/2020).






\appendix

\section{Spontaneous emission}\label{app:PL}
In this Section, we justify the neglect of spontaneous emission.
To do that, we include spontaneous emission (aka PL) in our electron distribution calculations by adopting Eqs.~(S3)-(S4) of Ref.~\cite{Sivan-Dubi-PL_I} for a single pair of initial and final electron energies, namely, 
\begin{align}\label{eq:rate_pEiEf}
\dfrac{\pi^2 V_\textrm{NP}^2}{12\hbar^2\varepsilon_h\varepsilon_0}|D(\e_i,\e_f,L)|^2 (\e_i-\e_f)\rho_e(\e_f) \rho_e(\e_i)(1-f(\e_f))f(\e_i)\rho_\text{phot}((\e_i-\e_f)/\hbar), 
\end{align}
where $V_\textrm{NP}$ is the NP volume (i.e., $\sim L^3$ for the current case) and $\rho_\text{phot}(\omega)$ is the photonic density of states. This expression gives the rate of photons emitted per unit electron energy $\e_i$ and per unit electron energy $\e_f$~\cite{photolum1} (hence, in units of sec$^{-1}$ Joule$^{-2}$) from an initial state $\e_i$ to the final state $\e_f$. 

\begin{figure}[h]
\centering
\includegraphics[width=\textwidth]{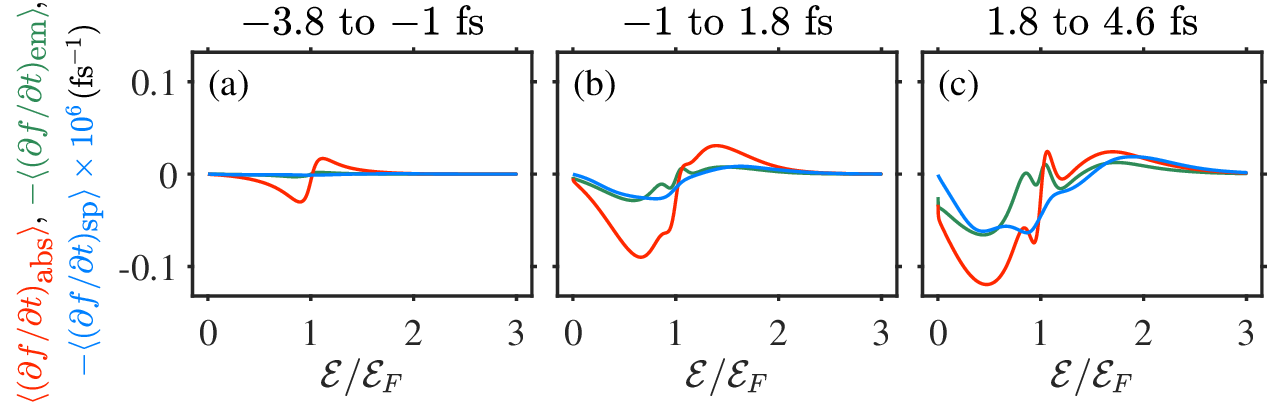}
\caption{\XYZ{(a)-(c) The same as Fig.~3(d)-(f) in~\cite{single_cycle_nlty_Letter}. The blue solid lines represent the local time average of the spontaneous emission part of the photon-e term Eq.~\eqref{eq:PL_in_out} for $E_0 = 1.2 \times 10^9$ V/m}{The local time average of the (stimulated) absorption (red), (negative of the) stimulated emission (green), and the spontaneous emission (Eq.~\eqref{eq:PL_in_out}, blue) parts of the photon-e terms for $E_0 = 1.2$ V/nm. The starting and ending points of the oscillation cycles are shown as titles.} (\XYZ{}{the spontaneous emission lines are} scaled by a factor of $10^6$ to allow for a better comparison with the stimulated emission and absorption parts) .}\label{fig:dfdt_abs_emi_sp_avg_1.2Vm-1_REF}
\end{figure}

The population at a specific energy state is affected by two types of processes. First, transitions between all initial states $\e_i$ to a fixed lower energy final state $\e_f$; then, there are transitions from a fixed initial state to all lower final energy states, similar to the red arrows in Fig.~\ref{fig:Du_abs_emi_out}(a). Subtracting these two types and integrating them appropriately, gives the change rate of population change at energy state $\e$ due to spontaneous emission events from all energy states $\e'$
\begin{multline}\label{eq:PL_in_out}
\left(\dfrac{\partial f(\e,t)}{\partial t}\right)_\textrm{photon-e,se} = - \dfrac{\pi^2 V_\textrm{NP}^2}{12 \hbar^2 \varepsilon_h \varepsilon_0}
\Bigg[\\
\int_0^\e d\e'|D(\e,\e',L)|^2 (\e-\e')\rho_(\e') (1 -f(\e',t))f(\e,t)\rho_\text{phot}((\e-\e')/\hbar)
\\
- \int_\e^\infty d\e'|D(\e',\e,L)|^2 (\e'-\e) \rho_(\e')(1 - f(\e,t))f(\e',t)\rho_\text{phot}((\e'-\e)/\hbar)
\Bigg].
\end{multline}
Note that unlike the calculations of Ref.~\cite{Sivan-Dubi-PL_I} where the integration was performed over all states separated by a specific emission photon energy (in order to determine the emission spectra), here, we integrate over all emission photon energies. 

Fig.~\ref{fig:dfdt_abs_emi_sp_avg_1.2Vm-1_REF} shows that the spontaneous emission is smaller than the stimulated emission and absorption by $\approx 10^6$ (see also Fig.~3 in~\cite{single_cycle_nlty_Letter}), thus providing justification for neglecting the spontaneous emission effects in this study.


%

\end{document}